\documentclass[letterpaper]{article}
\usepackage{natbib,alifexi,subcaption,units, color}
\usepackage{algorithm}
\usepackage{graphicx}				
\usepackage{hyperref}
\usepackage{todonotes}
\hypersetup{
    colorlinks = true,
    linkcolor = blue,
    filecolor = magenta,      
    urlcolor = cyan,
    citecolor = blue,
}
\usepackage{amssymb}
\usepackage{amsmath}
\usepackage{amsthm}
\usepackage{amsfonts}
\theoremstyle{plain}

\theoremstyle{definition}

\theoremstyle{remark}

\usepackage{algorithmic}

\title{Simulating short- and long-term evolutionary dynamics on rugged landscapes}
\author{
Leonardo Trujillo$^{1}$, Paul Banse$^{1}$ \and Guillaume Beslon$^{1}$\\
\mbox{}\\
$^1$Inria Beagle Team, Artificial Evolution and Computational Biology,\\
Universit\'e de Lyon, Inria, ECL, INSA Lyon, Universit\'e Claude Bernard Lyon 1, Univ. Lumi\`ere Lyon 2,\\
 CNRS, LIRIS UMR 5205, F-69622, France\\\
\{leonardo.trujillo, paul.banse, guillaume.beslon\}@inria.fr}
\begin{document}
\maketitle
\begin{abstract}
We propose a minimal model to simulate long waiting times followed by evolutionary bursts 
on rugged landscapes.
It combines point and inversions-like mutations as sources of genetic variation. 
The inversions are intended to simulate one of the main chromosomal rearrangements. 
Using the well-known family of NK fitness landscapes, 
we simulate random adaptive walks, i.e. successive mutational events constrained to incremental fitness selection. 
We report the emergence of different time scales: a short-term dynamics mainly driven by point mutations, followed by a long-term (stasis-like) waiting period until a new mutation arises. 
This new mutation is an inversion which can trigger 
a burst of successive point mutations, and then drives the system to new short-term increasing-fitness period. We analyse the effect of genes epistatic interactions on the evolutionary time scales.
We suggest that the present model mimics the process of evolutionary innovation and punctuated equilibrium. 
%
\end{abstract}
\section{Introduction}
A very simple model for artificial molecular evolution consists of genetic-like information stored 
and encoded  in a binary sequence. If this {\it digital genome} is subject to variations 
caused by mutations, along with selection, then a minimalist evolutionary dynamics can be 
simulated. Can such a simple model recreate different evolutionary time scales in a single framework?
This is a question that is not devoid of interest to evolutionary biologists, that has caused controversy when trying to understand how the causal relationship between micro and macro evolutionary processes could be~\citep{uyeda2011}. 
In this sense, in contrast to Darwinian gradualism, c.f. \cite[p. 32]{darwin1964}, the theory of punctuated equilibrium (PE) proposes that macroevolutionary dynamics consists of short periods of rapid evolution alternated by long periods of stasis with little evolutionary change, e.g.~\citep{gould1977}. 
However, some of the statistical methods used in the quantitative analysis of the historical data of macroevolutionary patterns, and their relationship with PE, have also been the subject of controversy and confusion~\citep{pennell2014}. 
On the other hand, several theoretical and computational models have addressed many features of PE. 
For example, two recent attempts to link the micro and the macro processes are:
the developmental gene hypothesis proposed by \cite{casanova2020}, that supports the genetic basis of PE; and the threshold response to increasing ecological pressures resulting from evolutionary diversification at an existing level of complexity, leading to punctuated expansions~\citep{ispolatov2019}.

Far from providing an ultimate answer to biologists, in this paper we report 
simulations that can simultaneously take into account changes across a range of well-differentiated 
time scales, without incorporating exogenous elements beyond the molecular evolution of our 
digital genome model (such as environmental contingencies). 
We use the~\cite{kauffman1993} NK model of molecular fitness landscapes, where 
$N$ is the length of the digital genome and $K$ is the ``epistatic'' coupling  between the $N$-sites
in the genome and determines the fitness landscape ruggedness~\citep{kauffman1987,kauffman1989,kauffman1993}.

In this  framework, we elucidate that 
the differences between the time scales are related to the  occurrence of chromosomal
rearrangements~\citep[Ch.~17]{griffiths2012} 
(at least those related to mutations that reverse the order of the genomic sequence, as reported in this work). 
Moreover, from the simulations outcomes we identify
{\it evolutionary bursts}. That is, once a new inversion promotes an increase in fitness -- after a long waiting time -- 
it triggers a successive occurrence of point mutations.
We show that in an early stage of the evolutionary process the system had exhausted the increasing-fitness point mutations  
around a local maximum of fitness. Then, once an inversion accesses a ``new'' domain of the fitness landscape, 
the exploration is re-started and is carried out mostly by point mutations.
Here, the meaning of ``new'' refers to domains of the fitness landscape that would be inaccessible through point 
mutations once a local maximum fitness value has been reached. 
This provides a very simple and evocative notion to think about evolutionary innovation and punctuated equilibrium. 

In the following section of this paper we introduce the model, with a description of our
double-stranded representation of binary genomes. Then we derive a new inversion-like
mutation operation, together with the standard bit-flip mutations. We continue with a
brief introduction of the NK model and the adaptive walk approach. Next, we present our simulations
and our main results. We close the paper 
with some conclusions suggesting that the present model mimics the process of evolutionary innovation and paves the way to simple simulations to explore dynamics such as punctuated equilibrium.

\section{The model}
\subsection{Genome representation}
We assume that the digital organism's genotype is a Boolean sequence of length $N\in\mathbb{N}$. 
In this way, all possible genome combinations define a $2^N$ Boolean hypercube space $\mathcal{Q}_N=\{0,1\}^N$.
Inspired by molecular biology, these genotypes are {\it double-stranded} binary vectors 
${\bf x}:=(x_1,\ldots,x_N)\in\mathcal{Q}_N$, 
with $x_i\in \{0,1\},\forall i \in \{1,\ldots, N\}$, where the {\it complementary} vector $\bar{\bf x}$ is defined 
such that $\forall x_i\in\{0,1\}$, $\bar{x}_i:= 1 - x_i$~, $\forall i\in\{ 1,\ldots,N\}$.
We assume that the sequences are circular, i.e.  periodic boundaries: 
$x_{N+i}=x_{i},\forall i \in \{1,\ldots, N\}$. From a biological point of view in this binary representation
each bit is a locus in a gene located on the first strand.  In this sense, our model mimics the molecular evolution of circular 
viruses~\citep{sole2018, tisza2020}.
For computational purposes, we model the molecular  evolution by selection dynamics as a two-step process. 
The first step involves the ``access to a new genotype'' by mutation, whereas the second step determines which 
mutated genotypes  will be selected (or fixed).
Let us remark that we do not include recombination, because we are modelling asexual replication. 

\subsection{Genetic mutation operations}
Besides classical single locus mutations (bit-flip), chromosomal inversions contribute to another level of
genomic variation, which consists of an order reversal in a segment of the genome. 
In this case the double-stranded characteristic of DNA plays an important role because
a molecular inversion is not a mere permutation between sites of the sequence.
Rather, it consists in a permutation of the complementary strand, to then be exchanged 
with the main strand within the segment where the inversion occurs~\citep[p.~614]{griffiths2012}. We
model the chromosomal-like inversion as a two step procedure: a {\it conjugation operation}, followed by
a {\it permutation operation}.

We define the conjugation operation $\mathrm{\hat{C}}$ such that 
\begin{equation}\label{Eq:C}
\mathrm{\hat{C}}:(x_i,x_{i+1},\ldots,x_{j-1},x_j)\longrightarrow(\bar{x}_i,\bar{x}_{i+1},\ldots,\bar{x}_{j-1},\bar{x}_j),
\nonumber 
\end{equation}
for $i,j\in \{1,...,N\}$. Note that, as the genome is circular, there is no relation of order between $i$ and $j$ (i.e. $\mathrm{\hat{C}}$ is well-defined even when $i>j$).
The permutation operation $\mathrm{\hat{P}}$ is defined as
\begin{equation}\label{Eq:P}
\mathrm{\hat{P}}:(x_i,x_{i+1},\ldots,x_{j-1},x_j)\longrightarrow (x_j,x_{j-1},\ldots,x_{i+1},x_i),
\nonumber
\end{equation}
for $i,j\in \{1,...,N\}$. Then, we have the 
(two-step) inversion operation:
\begin{equation}
\mathrm{\hat{I}}\equiv\mathrm{\hat{C}}\circ\mathrm{\hat{P}}.
\label{Eq:Inv}
\end{equation}
Trivially, it can be verified that $\mathrm{\hat{C}}$ and $\mathrm{\hat{P}}$ commutes:
$\mathrm{\hat{C}}\circ\mathrm{\hat{P}}=\mathrm{\hat{P}}\circ\mathrm{\hat{C}}$.
Besides, Eq.~\ref{Eq:Inv} also defines a point mutation, i.e. when $i=j$, then $\mathrm{\hat{I}}$ is a single bit-flip, equivalent to
the biological transversion-type point mutation~\citep[p.~555]{griffiths2012}. 
These operations can be easily translated into a model of computation
capable of simulating the evolutionary dynamics of a genetic-like Boolean sequence
of length $N$. 
%

\subsection{NK fitness landscape}
After the picturesque notion of  “fitness landscape”, introduced in evolutionary biology 
by \cite{wright1932}, adaptive evolution is often conceptualised as hill-climbing walks on rugged 
landscapes. This metaphor even became a ``tool'' to quantify (and think) many features 
about molecular evolution~\citep{gavrilets2004,obolski2018}. 
A well tailored model to simulate artificial molecular evolution is the 
 NK fitness landscape~\cite[Ch.~2]{kauffman1993}.

In the NK model, besides the genome length $N\in\mathbb{N}$, there is an integer  
$K\in\mathbb{Z}_{(0,N-1)}$, which describes the epistatic interaction
between the sites in the genome and the contribution of each component to the total fitness, 
which depends on its own value as well as the values of $K$ other sites. 
The fitness per bit is formally defined as:
\begin{equation}
f_i:\{ 0,1\}^{K+1}\longrightarrow [0,1), \,\,\,\, \forall 1\leq i \leq N.
\nonumber
\end{equation}
Here $f_i(x_i,x_{i_1},\ldots,x_{i_K})$ 
depends on the state of site $x_i\in\{0,1\}$ and $K$ other sites $x_{i_K}\in\{0,1\}$. The $f_i$'s are given by $N\cdot2^{K+1}$ independent and identically distributed random variables sampled from a given uniform probability distribution. Different features of the NK model with more general distribution functions have been analysed by, for example, \cite{evans2002}, \cite{durrett2003} and \cite{limic2004}.

The two frequently used neighborhood models are  full disordered  and  $K$-regular graphs:
(i) The random neighborhood model, where $i$ and $K-1$ other sites are chosen at random according to a
uniform distribution from $\{1,2,\ldots,N\}$;  
(ii) The adjacent neighborhood model, where $i$ and $K$ other sites are successively ordered, i.e. 
$i,i+1,\ldots,i+K$ (each variable modulo $N$ when using periodic boundary conditions). 
The total fitness $f\in[0,1)$ for the sequence ${\bf x}\in\mathcal{Q}_N$ is then defined as:
\begin{equation}
f({\bf x}):=\frac{1}{N}\sum_{i=1}^{N}f_i(x_i,x_{i_1},\ldots,x_{i_K}),
\label{Eq:NK}
\end{equation}
where $\{ i_1,\ldots,i_K\}\subset \{1,\ldots,i-1,i+1,\ldots,N \}$. 

The most important feature of the NK model is that the parameter $K$ ``tunes'' the landscape ruggedness, that is the distribution of fitness local maximums, ranging from non-epistatic interactions when $K=0$ (a Mount-Fuji-like landscape with a single peak), to the full rugged
(or random) landscape when $K=N-1$. This allows us to analyse the time scales as a function of the ``disorder'' of local maximums (fitness peaks) over the landscape.

\subsection{Adaptive walks}
The evolutionary dynamics of a single digital organism can easily be translated  
in terms of adaptive walks over the fitness landscape. That is, once given
a starting sequence ${\bf x}\in\mathcal{Q}_N$, it varies through successive 
mutations (one bit-flip or one inversion per time step) resulting in a mutated sequence
${\bf y}\in\mathcal{Q}_N$. Then, the fitness $f({\bf x})$ and $f({\bf y})$ are calculated according to the NK model (\ref{Eq:NK}). 
If $f({\bf y})>f({\bf x})$, the mutated sequence is selected, otherwise other mutations on
${\bf x}$ are tested
until the fitness increases. Therefore, fitness decreasing 
mutations cannot be fixed, analogous to
the ``strong selection, weak mutation'' regime in population genetics~\citep{gerrish1998}.
The consequence of this constraint is that once a local maximum of fitness is reached, no point mutation will allow the evolutionary dynamics to continue. It is just stuck in a local peak. Unless a beneficial inversion-type mutation is involved, and possibly opens a new way to increase the fitness.
This would be equivalent to ``escape'' from the local peak. 
In this new version of adaptive walks on NK landscapes, we include the probability
$p\in[0,1]$ associated with the occurrence of inversion-like mutations between two sites $i$ and $j$ of the genome, such that $i\neq j\in\{1,\ldots,N\}$ (and the probability of point mutations is $q=1-p, \therefore i=j$). When $p=1$, all the mutations are inversions (without point mutational events, i.e. $q=0$), and vice versa. If $0< p < 1/2$, the inversions are less frequent than the point mutational events. Then, the mutation interval $[i,j]\in\{1,\ldots,N\}$ is  drawn from a pseudo random number generator. 
Each time-step is determined by a  mutational (random) event constrained to an incremental fitness selection.
With these computational recipes we can simulate
successive mutated genotypes and their fitness values,  to generate 
fitness time series of adaptive evolutionary ``hikes'' for different instances of the fitness landscapes and different initial genomes. 
\subsubsection{Experimental setup}
\begin{figure}[ht]
\begin{center}
\includegraphics[width=0.95\linewidth]{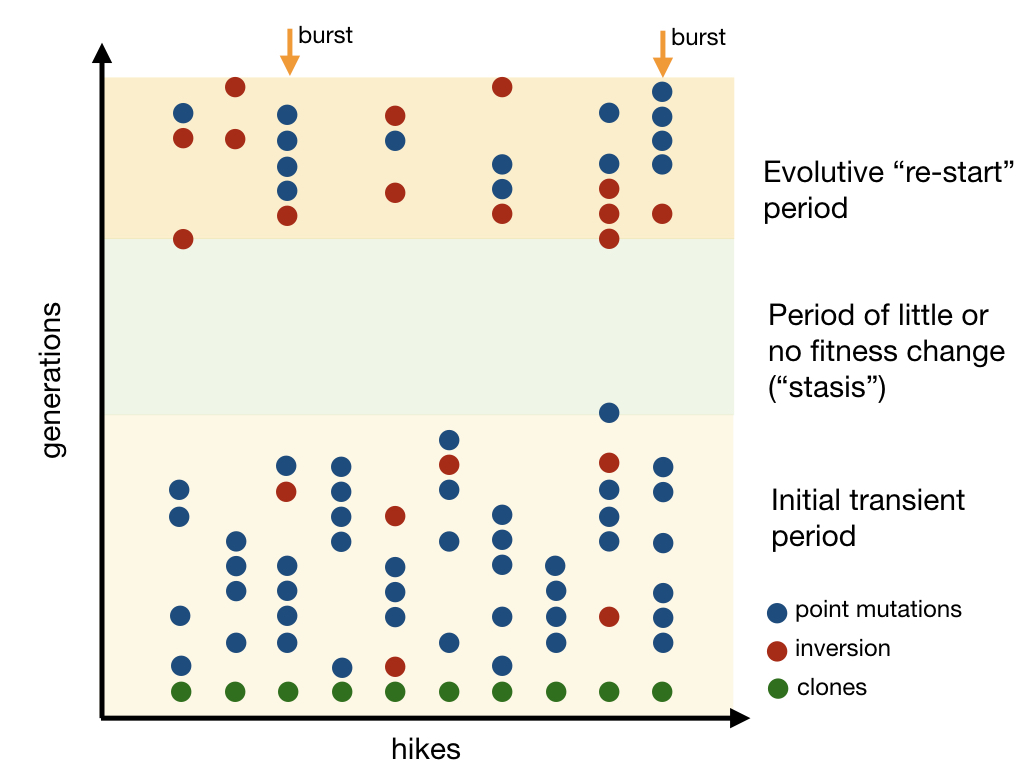}
\caption{{\bf Hikes schematic.} 
The vertical axis represents generations (i.e. evolution time-steps).
Each point indicates when a mutation is fixed. 
Green dots represent ``clones'' of a given initial genome ${\bf x}_0$. 
Blue and red dots symbolise point and inversion mutations respectively. 
Three time scales are outlined. An initial transient period in which point mutations predominate. %
Followed by a ``stasis''-like period, where the occurrence of mutations is very low or null. 
Finally, a period that ``re-initiates'' the evolutionary dynamics, which is only possible
if mutation by inversion occurs. The idea of evolutionary burst is outlined, which is roughly a 
succession of point mutations after an inversion occurs.
}
\label{fig:schematic}
\end{center}
\end{figure}
To elucidate the relationship between the evolutionary time scales and the two types of mutations considered in our model, we implemented a series of repetitions of the adaptive walk described above to create a pool of trajectories over the fitness landscape. That is, from a given initial genome ${\bf x}_0\in\mathcal{Q}_N$ of length $N$ we run different instances of the evolutionary process up to maximum number of time steps $\tau_{max}$. Theses instances are like a branching random walk, but instead of several descendants from a given genome in each generation (like in population genetics), we generate $2N$-hikes starting at the same ${\bf x}_0$ (imitating independent evolutionary experiments with ``clones'' of a same initial genome ${\bf x}_0$). Therefore, each hike is an independent exploration of accessible paths over a (sub)domain of the fitness landscape. These samples are {\it quenched} in the sense that we do not run simulations of trajectories for a given a set $\{ {\bf x}_0 \}\in\mathcal{Q}_N$ of independent initial genomes, but, as mentioned above, each time-step is driven by a  mutational (random) event. They are composed by two probabilistic processes, the first one is a random choice of the type of mutation (bit-flip or inversion) accordingly to the probability $p$; in the second one, the sites to be mutated are raffled. For example, this can by done seeding a pseudo random numbers generator independently for each process. See figure~\ref{fig:schematic} for a schematic of the time evolution of hikes.
\section{Results}
\begin{figure}[h]
\begin{center}
\includegraphics[width=0.495\linewidth]{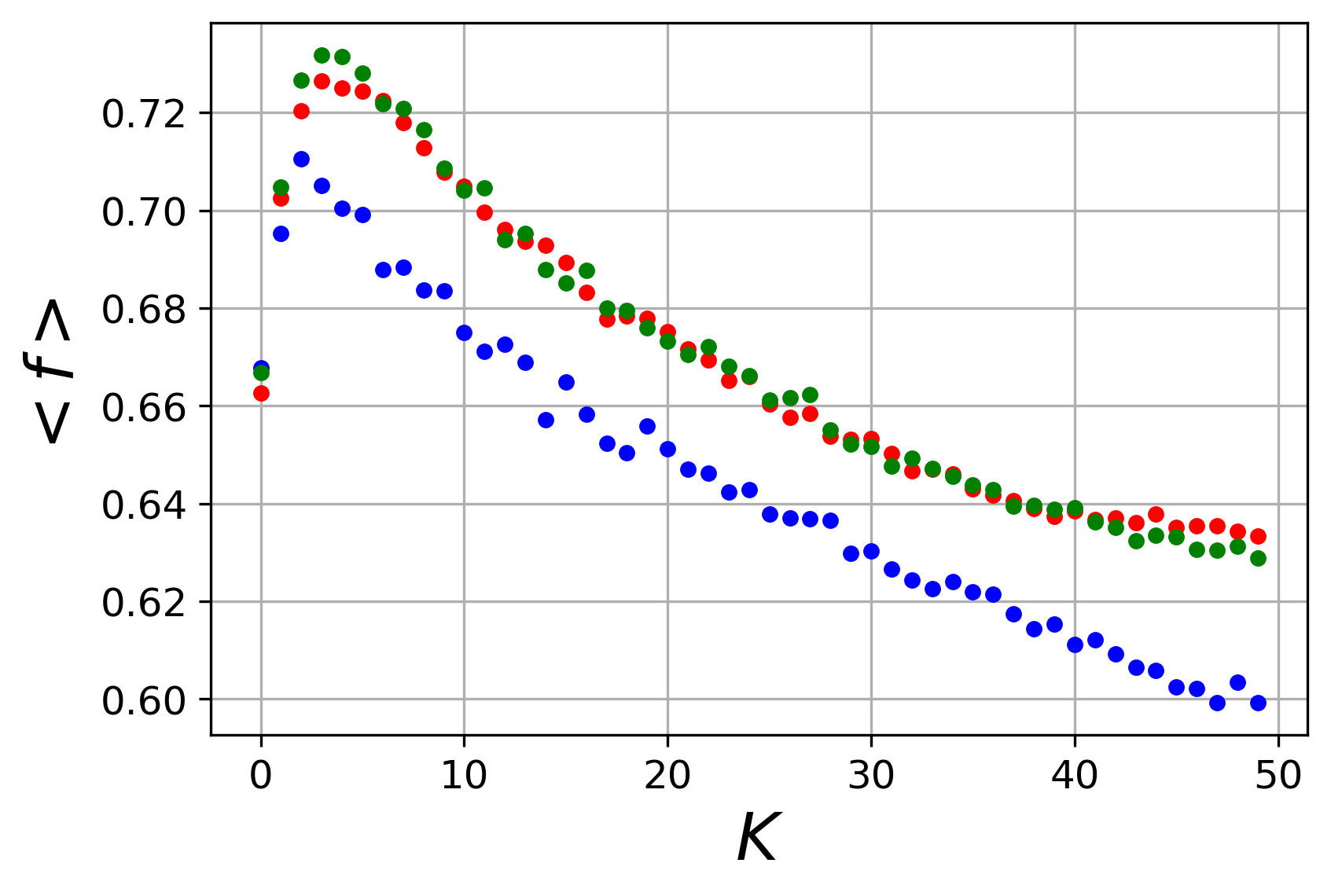}
\includegraphics[width=0.495\linewidth]{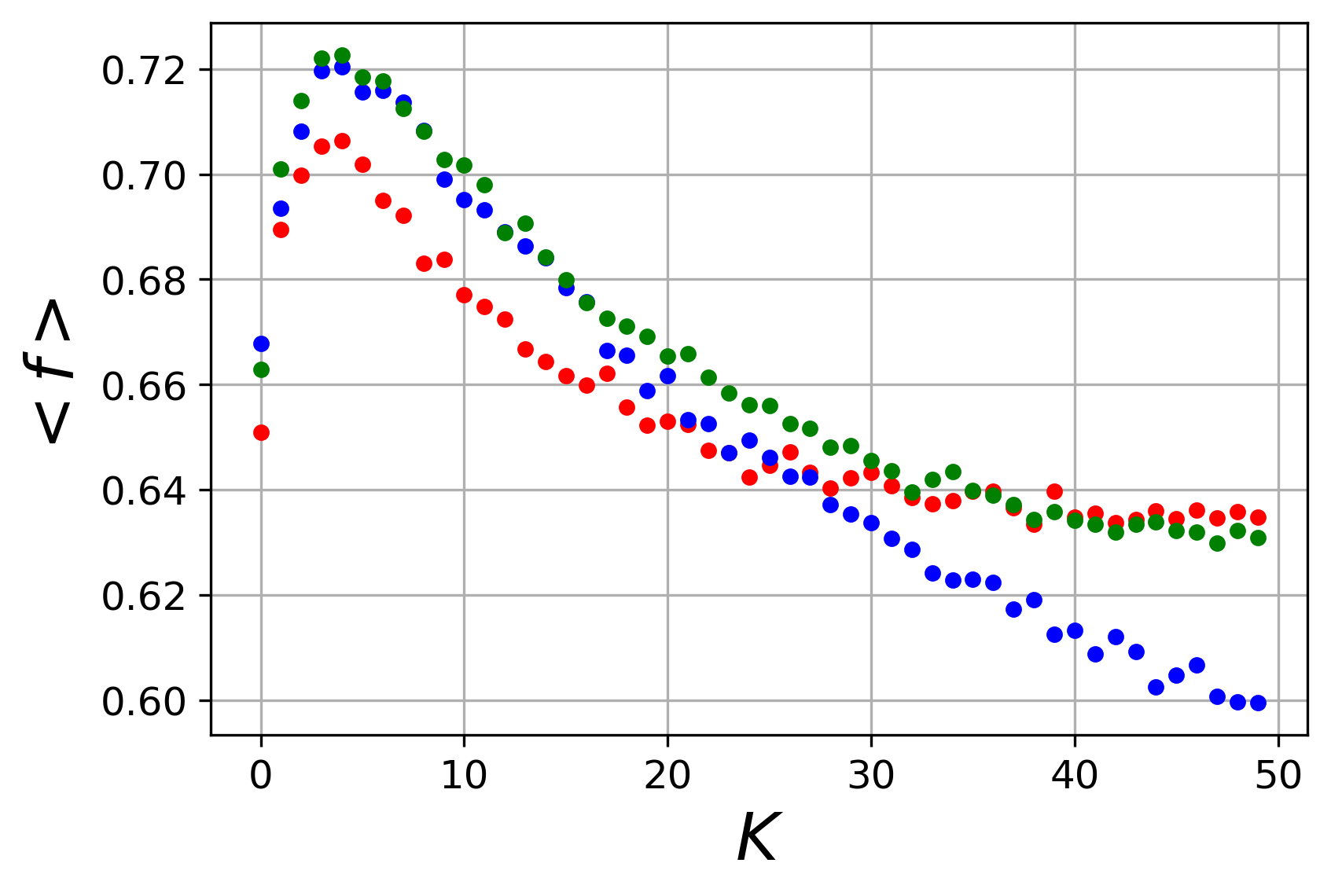}
\caption{{\bf Fitness benchmark.} Expected fitness value for
genomes of length $N=50$ versus the epistatic parameter $K=(0,\ldots,49)$.
Blue dots correspond to point mutations only ($p=0$)  while the red ones correspond to inversions only ($p=1$). 
Green dots are for equiprobable occurrence of inversions and point mutations ($p=1/2$).
{\bf Left:} adjacent epistatic interactions. {\bf Right:} random epistatic interactions. Each point is the average for $100$ independent hikes.
}
\label{fig:finalFitness}
\end{center}
\end{figure}

Now we summarise the results of our empirical exploration. 
We display illustrative results\footnote{Data and
code are available at \url{https://gitlab.inria.fr/letrujil/NK_hikes}.} for genomes of length $N=50$.
\paragraph{Fitness benchmark} First, before exploring the time scales, we gauged the fitness values for different instances of
NK landscapes, in function of $K$ for adjacent and random neighbourhoods.    
We analysed three mutational regimes: inversions and point mutations only ($p=1$ and $p=0$ respectively), and the mixing of both operations ($p=1/2$).
In figure~\ref{fig:finalFitness} we can verify that  when $K=0$, the expected fitness value is
$\langle f \rangle\simeq 2/3$, and when $K$ increases, then the expected fitness value decreases (for point mutations $\langle f \rangle \rightarrow 0.60$ and for inversions
$\langle f \rangle \rightarrow 0.63$). The trends displayed in figure~\ref{fig:finalFitness} are congruent with the well known ``complexity catastrophe'' in the NK model, i.e.  the tendency of (averaged) local fitness peaks to decrease when the epistatic interactions increase~\citep[p. 52]{kauffman1993}. This trend depends on the probability distribution for the contributions to fitness $f_i$ (which in our case is an uniform distribution of identically distributed random variables)~\citep{solow1999}.

Figure~\ref{fig:finalFitness} also shows that when the system evolves under the sole inversion regime,  fitness values are generally higher than
in the point mutation regime. However, this phenomenon is stronger in the random neighbourhood model: When the epistatic interactions
are fully random, the fitness values in the point mutation regime are higher than those of 
the inversion regime below $K\simeq25$. This is likely to be due to the fact that inversions are local operators, hence being more efficient when epistatic interactions are local. In the following of the paper all simulations and analyses use the adjacent epistatic neighbourhood.

\subsection{Inversions trigger evolutionary bursts}

\begin{figure*}[!htp]
\begin{center}
\includegraphics[width=0.3\linewidth]{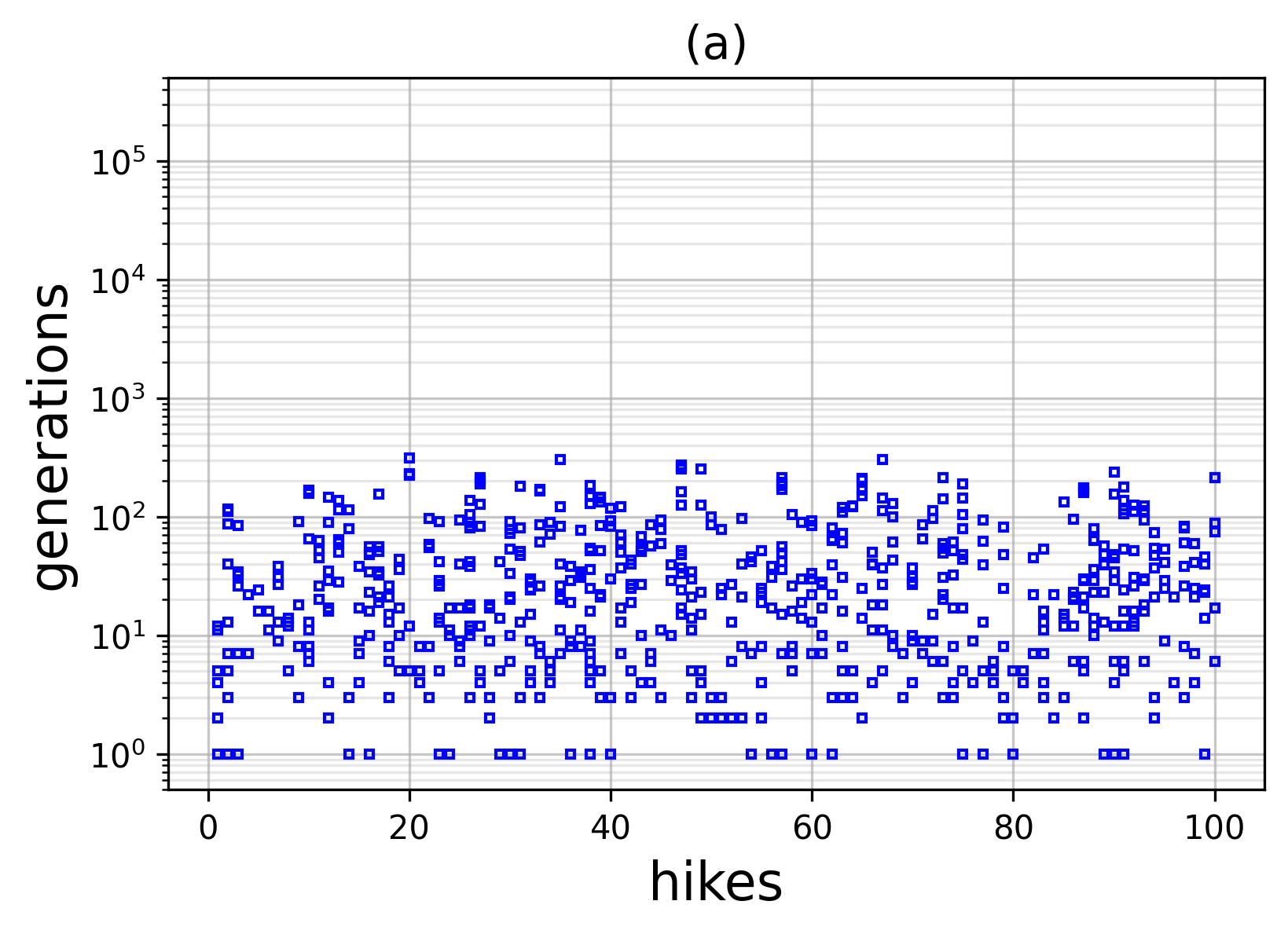}
\includegraphics[width=0.285\linewidth]{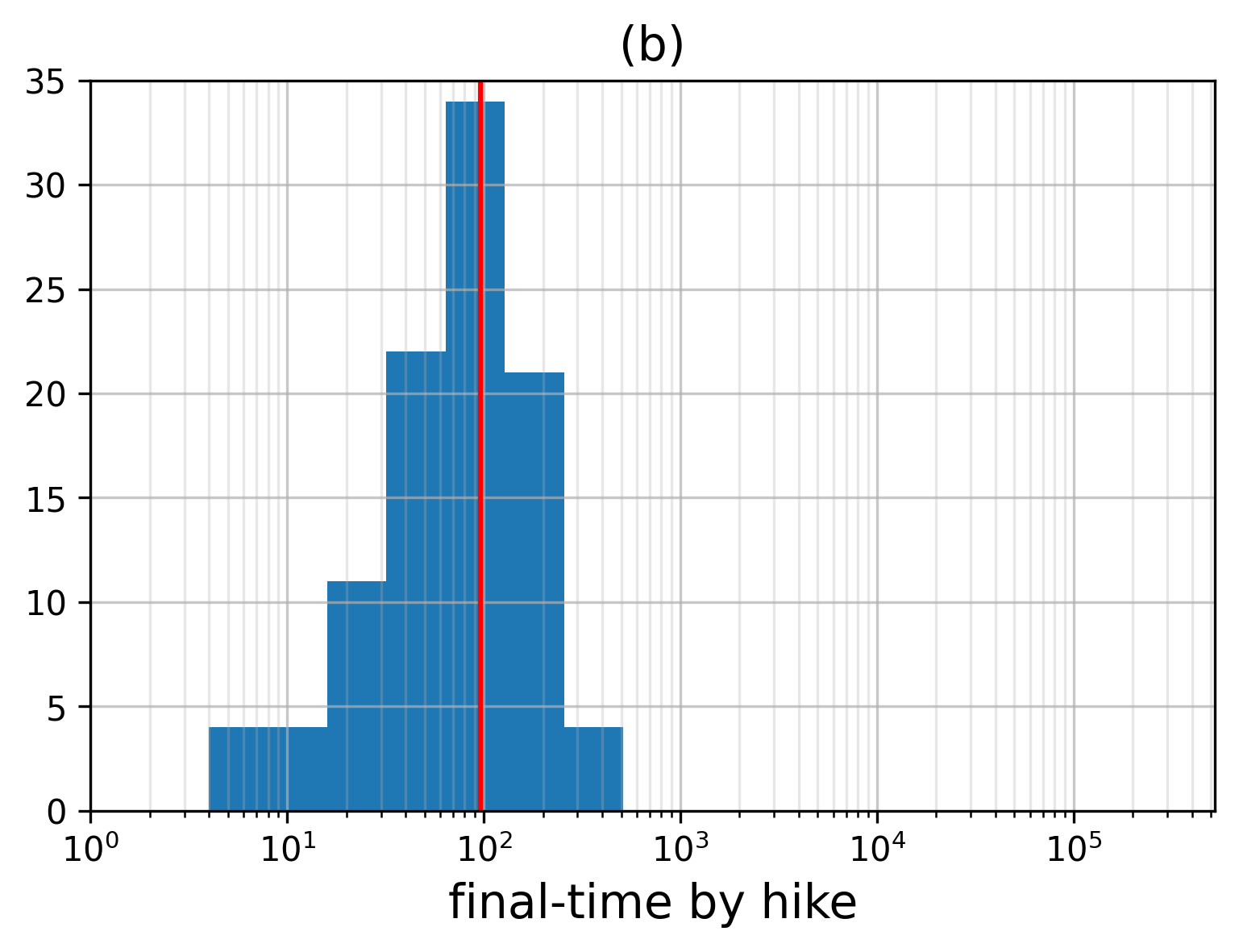}
\includegraphics[width=0.29\linewidth]{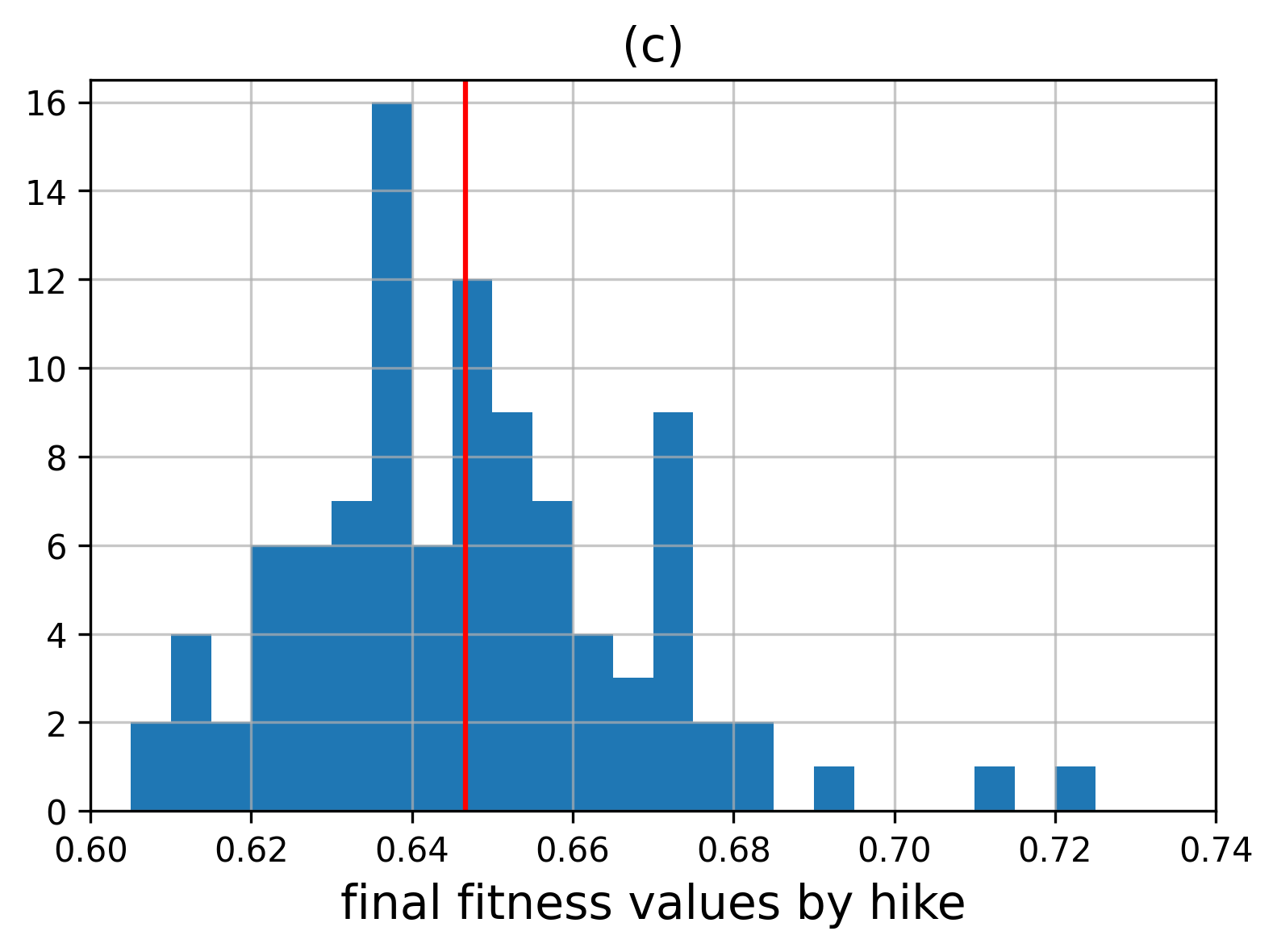}
\includegraphics[width=0.3\linewidth]{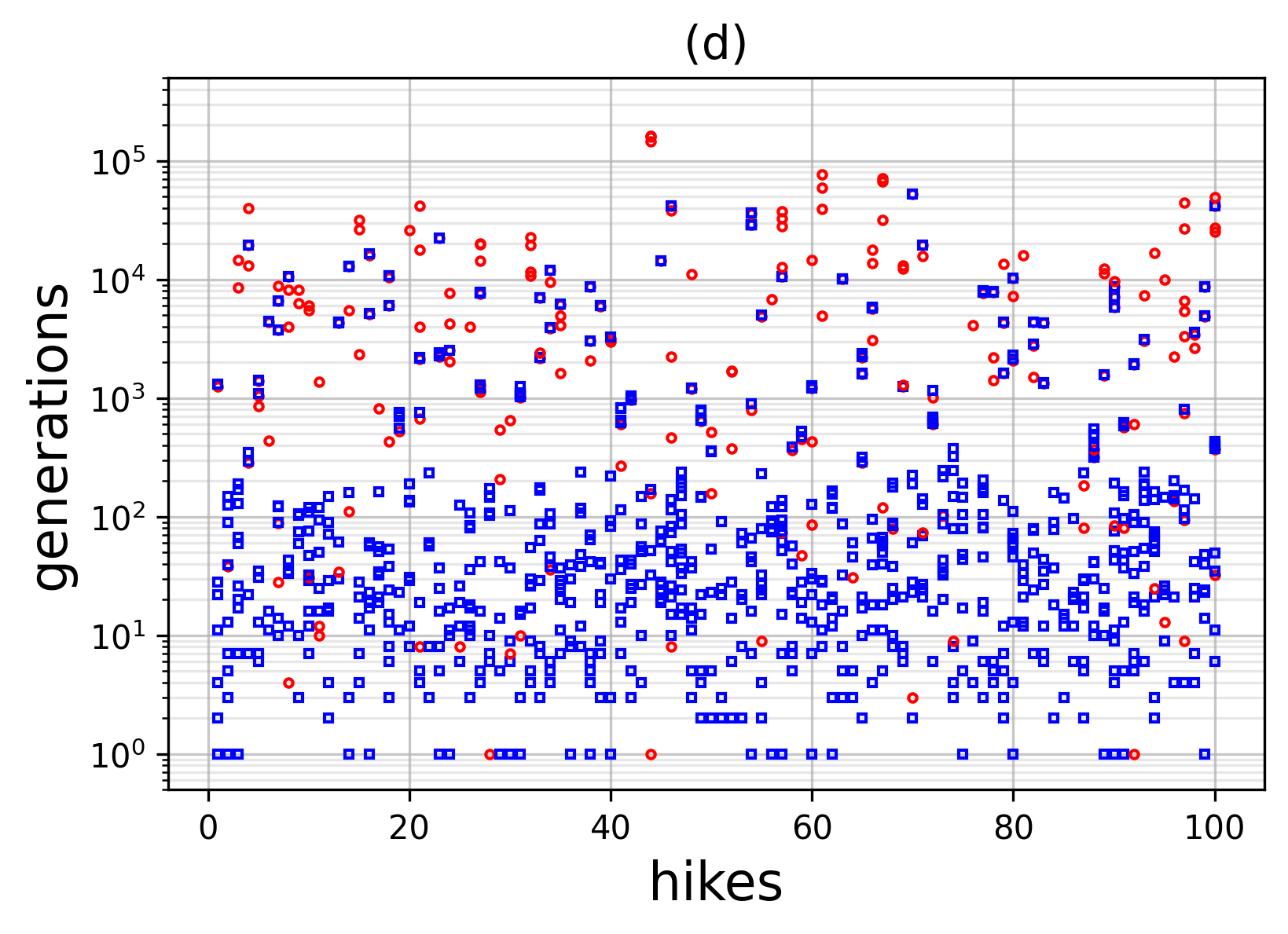}
\includegraphics[width=0.287\linewidth]{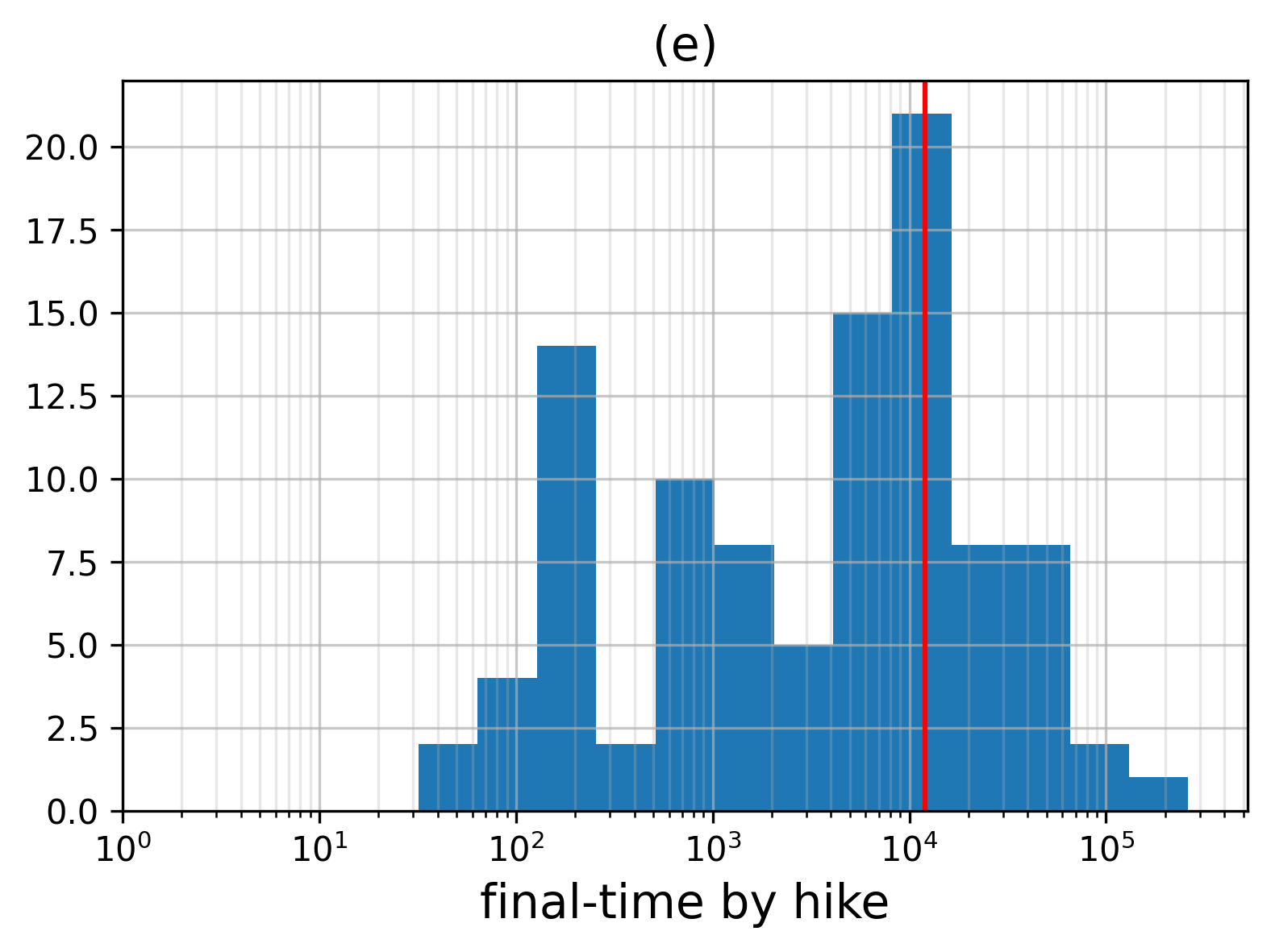}
\includegraphics[width=0.29\linewidth]{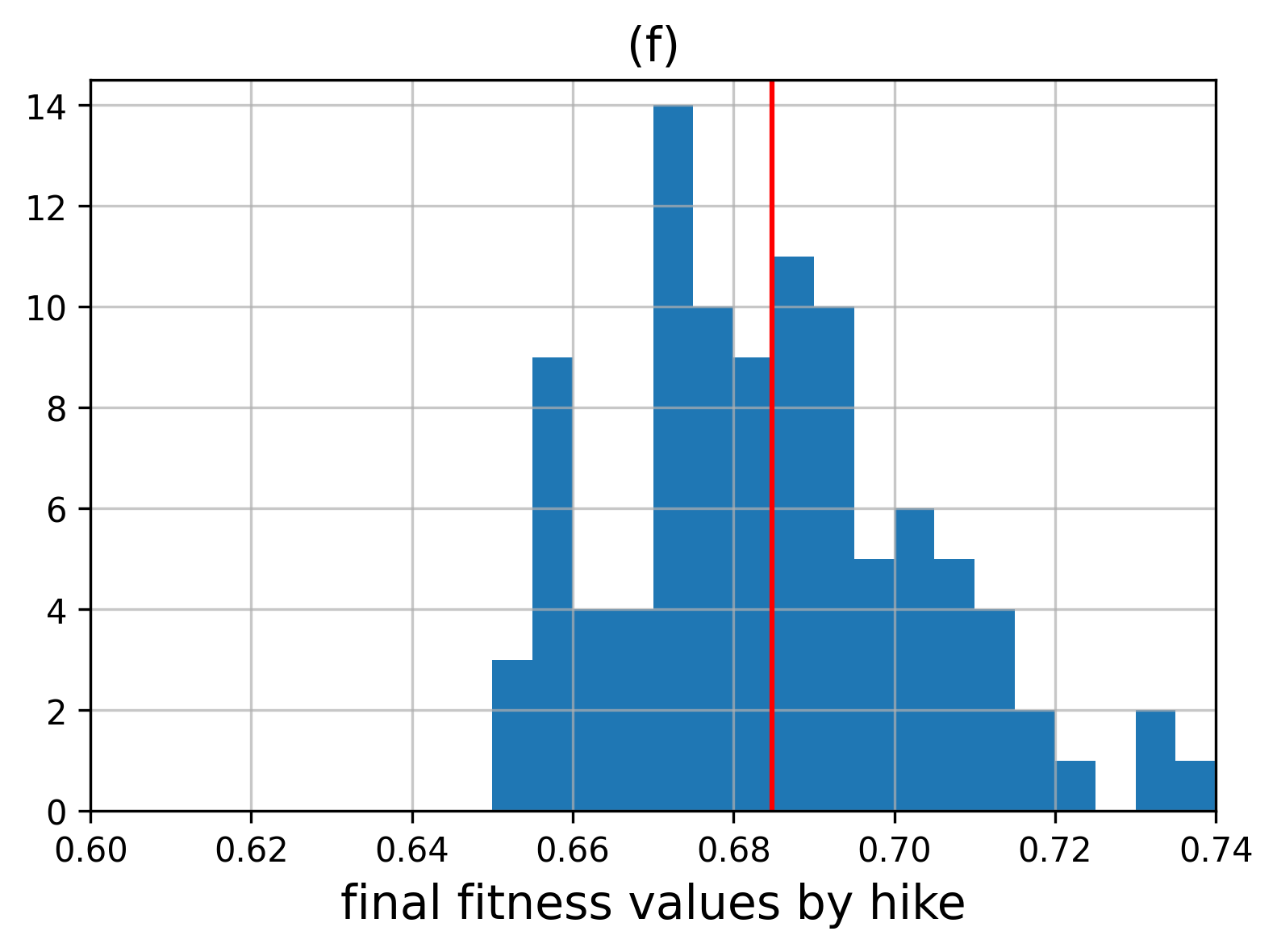}
\caption{{\bf Effect of inversions on the evolutionary dynamics} for $N = 50$, $K=20$, 
$\tau_{max}=2\times 10^5$ generations for a given
initial genome ${\bf x}_0$. 
{\bf Top:} $p=0$ (bit-flip only); {\bf Bottom:} $p=0.1$ ($10$\% of the mutational events are inversions). (a) and (d) are scatter plots of the mutational events per hike and for the two mutational conditions (blue open square: bit-flips; red open circles: inversions). The histograms (b) and (e) count the distribution of the final time step (i.e. generation of the last fixed mutational event, being it a bit-flip or an inversion) for the two mutational conditions. (c) and (f) display the distribution of the final fitness values at generation $2\times 10^5$ for the two mutational conditions. The red lines indicate the respective mean values.  
}
\label{fig:hikes1}
\end{center}
\end{figure*}

\begin{figure*}[!hbp]
\begin{center}
\includegraphics[width=0.85\linewidth]{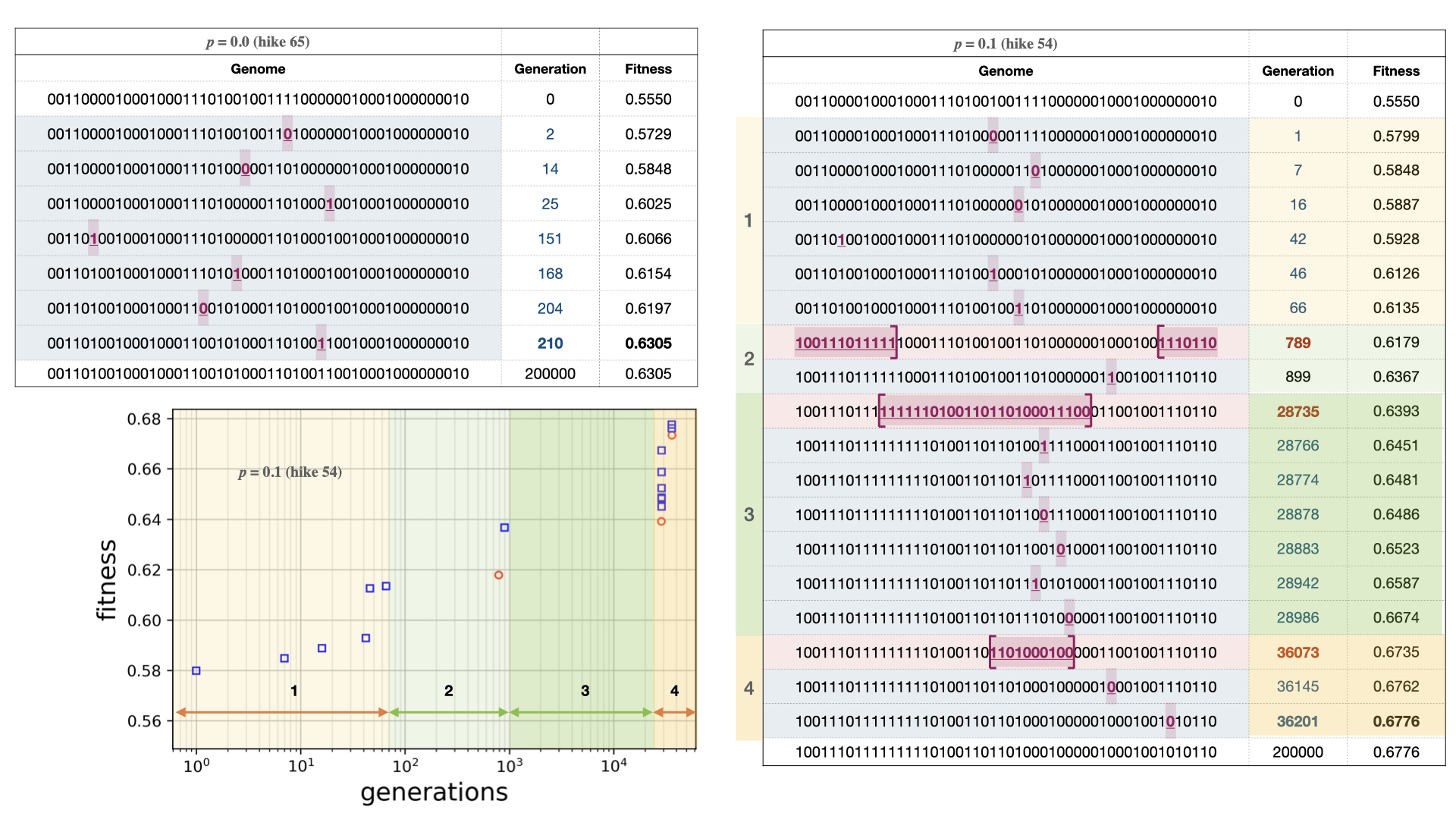}
\caption{{\bf Fitness bursts.} Fitness time-series of two representative hikes for
$N=50$, $K=20$ and $\tau_{max}=2\times 10^5$. The pale blue rows emphasise the genomes muted by a bit-flip (highlighted with pale pink), while the pale pink rows emphasise inversion mutational events. The underlined mutated segments in the genome are framed by brackets, and the bits are boldfaced. 
{\bf Top left panel:} shows the generations when the mutational events were fixed for hike 65, when $p=0.0$. 
{\bf Right panel:} illustrates the case when $p=0.1$ for hike 54, showing the stasis-like period and the occurrence of the evolutionary bursts. 
{\bf Bottom left figure:} Displays the fitness time-series of hike 54.
See main text, and figure \ref{fig:schematic}, for the description of the $4$ evolutionary phases. 
}
\label{fig:tables}
\end{center}
\end{figure*}

\paragraph{The access to new fitness peaks} Following the experimental protocol described above, we evaluate $100$ hikes initialized with a same random genome $x_0$ for  $\tau_{max}=2\times10^5$, $N=50$
and $K=20$. Figure~\ref{fig:hikes1}~(a) shows a benchmark of the evolutionary process with point mutations
alone. The total final time per hike (i.e. the time when the last mutation has been fixed) doesn't even reach $10^3$ generations, verifying that the hikes 
stop at local fitness peaks (i.e. from $\sim 10^2$ generations up to $\tau_{max}$ nothing else happens). 
Moreover, the elapsed times between mutations are generally of some tens of generations
(see for example the top left panel in fig.~\ref{fig:tables}).
When inversions are allowed (i.e. $p\neq 0$, figure ~\ref{fig:hikes1}~(d)) we can verify that
the evolutionary dynamics lasts much longer than when $p = 0$. This can be appreciated in figure~\ref{fig:hikes1}~(b, e).
Inspecting this figure, we can notice that there are hikes that, like the previous case, stay blocked in 
a local maximum of fitness. However, and it is the key finding of the present work, there exist late 
mutational events --driven by inversion mutations-- allowing the evolutionary process to continue far away on time.
This can be better appreciated in figure~\ref{fig:hikes1}~(b, e), where the tails of the distribution of the final time per hike are larger when $p\neq0$ than when $p=0$.
The total number of generations when inversions are fixed can be almost $30$ times longer than the total number of generations reached when point mutations are the only source of genome variation in the evolutionary dynamics.
Moreover, the action of inversions can drive the dynamics to higher fitness values. This is shown in
figure~\ref{fig:hikes1}~(f), where the maximum value of fitness in the case $p\neq 0$ is higher than the case $p=0$ (figure~\ref{fig:hikes1}~(c)).
This shows that some ``new'' fitness peaks are ``discovered'' in comparison with the hikes when $p=0$. In the histogram this corresponds
to the bins of fitness between $0.68$ and $0.73$. So we can paraphrase that not only can evolution ``go further'', it can also discover something new: Peaks that were previously inaccessible over the fitness landscape, are now accessible thanks to the action of inversions that provoke large-scale reorganisations of the genome of the digital organism simulated in this work. 

\paragraph{An explicit example of stasis and bursts} In figure \ref{fig:tables} (bottom left and right panels)
we display an example from the pool of hikes when $p=0.1$ (for hike $54$ in figure~\ref{fig:hikes1}~(d)). It can be seen that the exploration of the fitness landscape is initially carried out by point mutations until  the sixty-sixth generation (this time-domain is labeled as $1$ in the figure). Then the dynamics remain in an invariant state for a relatively long period of time that lasts until the time-step $789$ (time-domain $2$). At that time an inversion is fixed which is rapidly followed by the fixation of a point mutation at generation $899$. Then, the evolutionary process returns to a ``stasis-state'' for a long period (between generations $899$ and $28,735$ -- time-domain $3$). Finally, at generation $28,735$ an inversion occurs that boosts evolution again (time-domain $4$). A very interesting observation is that at this stage there are a series of successive point mutations (similar to the number of point mutations that occurred at the beginning of evolutionary dynamics). We call this sudden mutational events an {\it evolutionary burst}.
This can be interpreted, using the fitness landscape metaphor, as if the system has ``jumped'' from a local fitness peak to a new one, hence discovering a new domain of the landscape.  
Then, the system begins to explore and climb this new peak until it reaches a local (with respect to Hamming distance) optimum of fitness. 
In this example, once again a new inversion occurs (at generation $36,073$) and the fitness keeps increasing again for a while. As we can see in  figure~\ref{fig:tables}, bottom-left panel, if we compare with the early stages
of the evolutionary process (time-domain $1$), the gain of fitness is larger on domain $4$ than on the first one where only point mutations were fixed. This shows that, when an inversion take place, it can boost evolution.

\subsection{Effect of the epistasis degree {\it K}}

\begin{figure*}[!htp]
\begin{center}
\includegraphics[width=0.3 \linewidth]{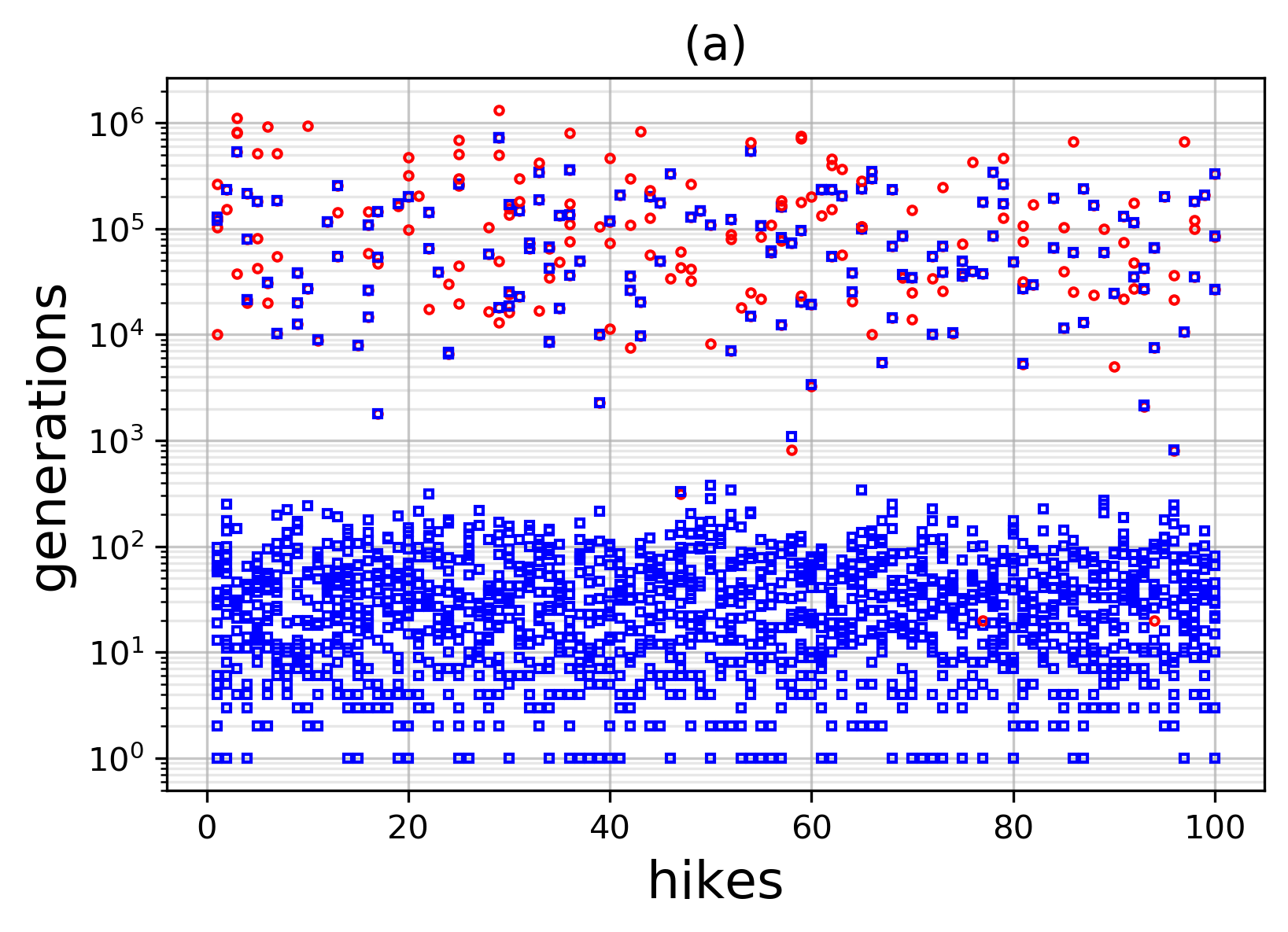}
\includegraphics[width=0.3\linewidth]{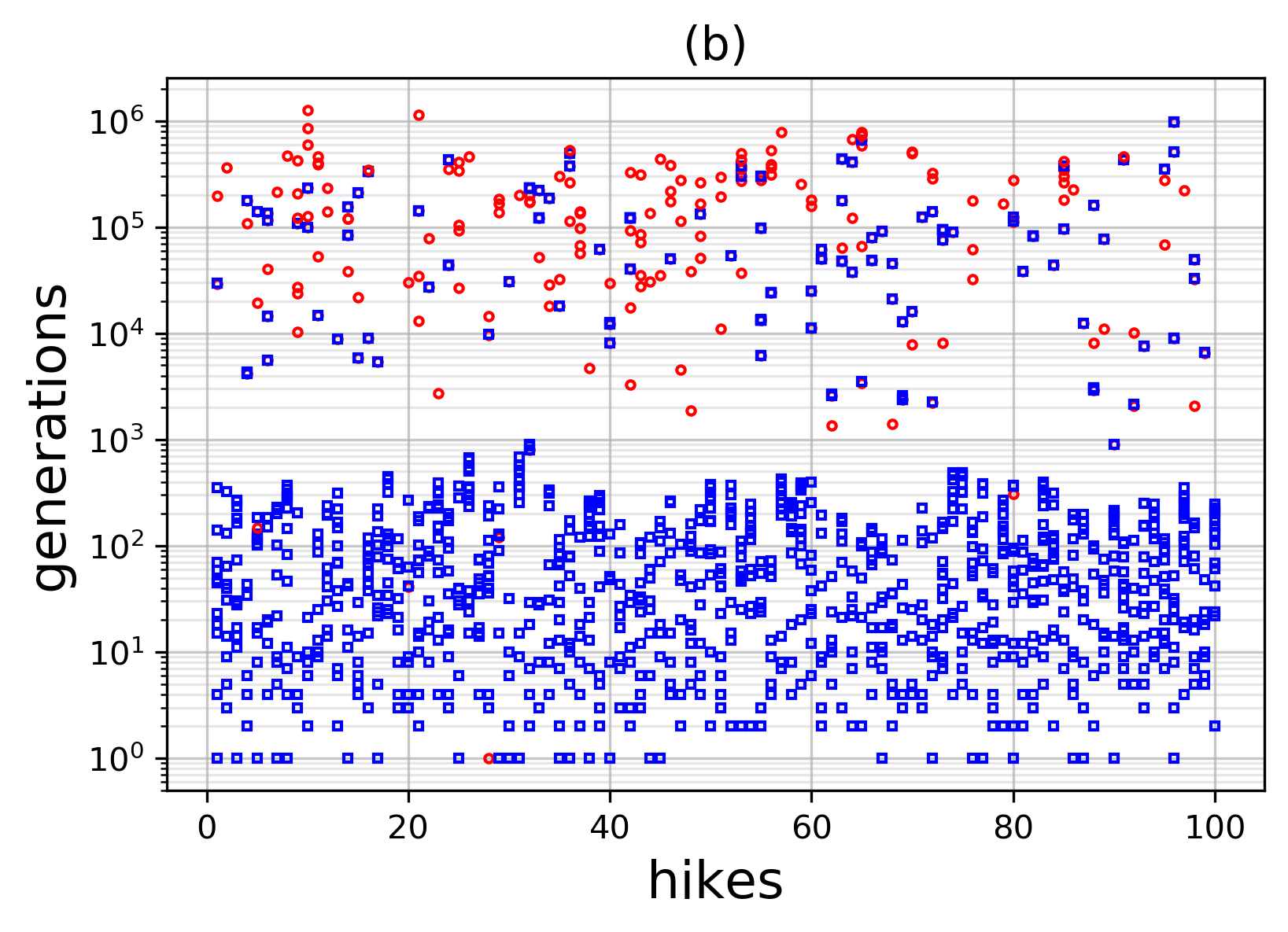}
\includegraphics[width=0.3\linewidth]{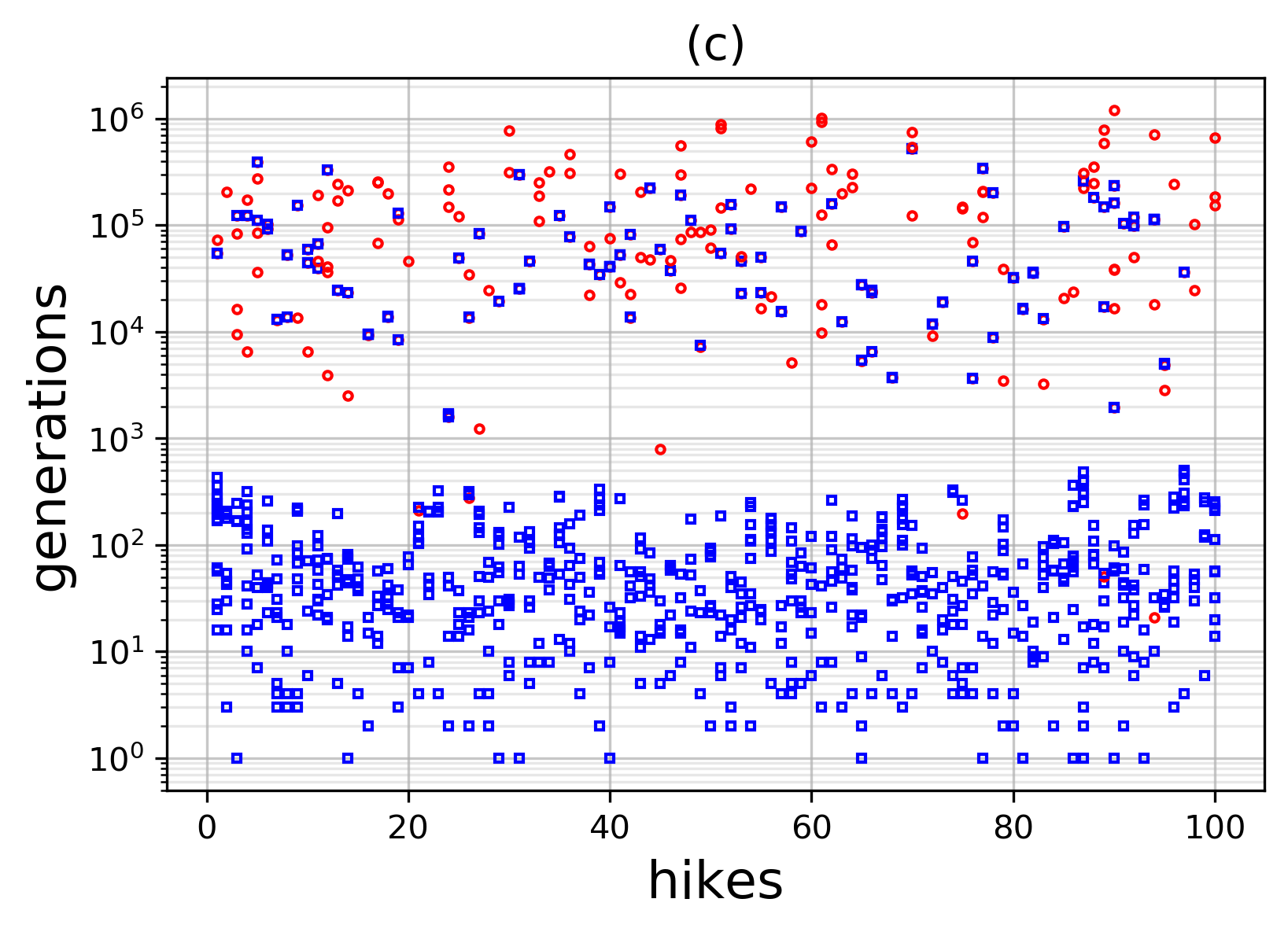}
\includegraphics[width=0.3\linewidth]{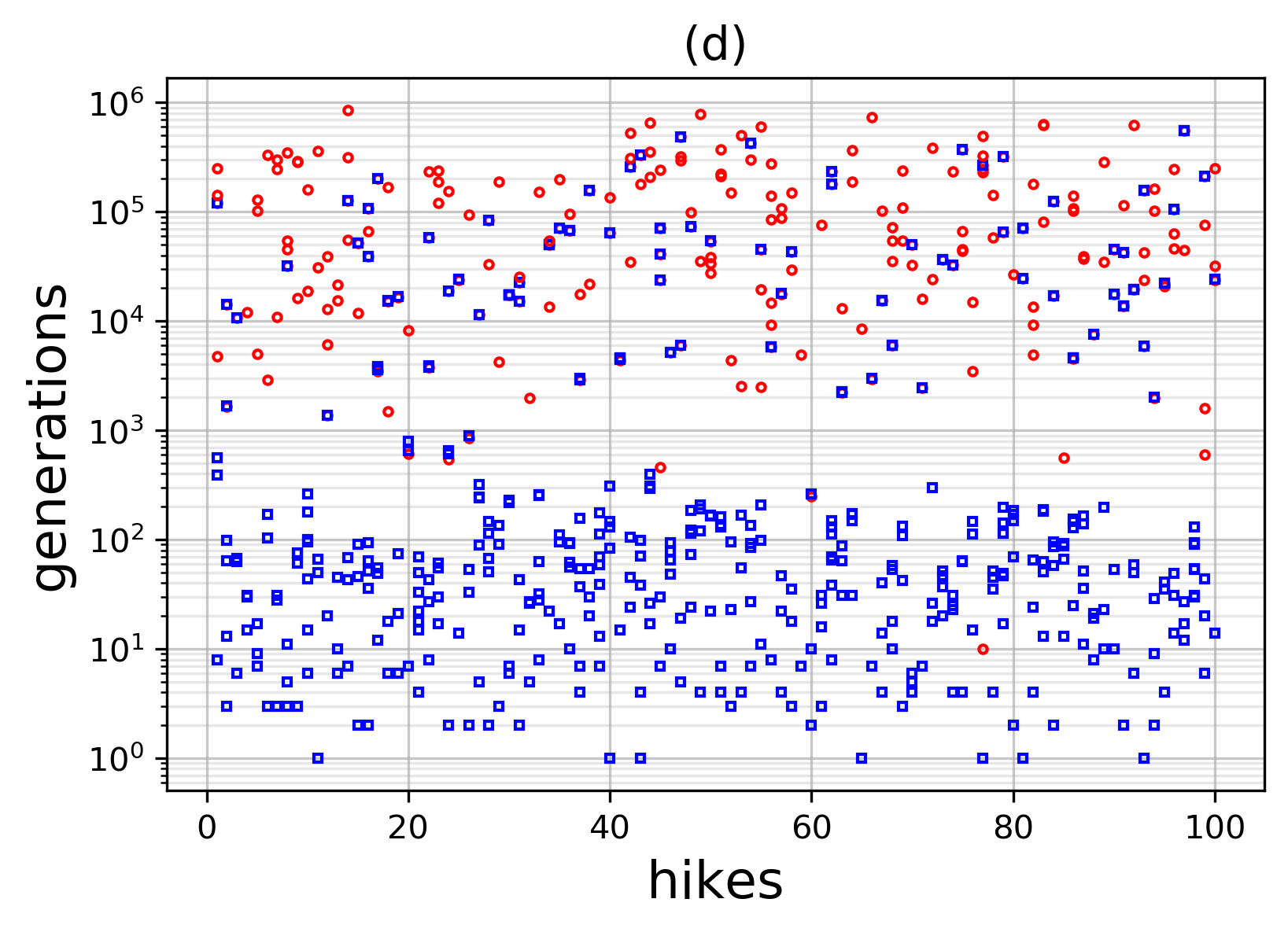}
\includegraphics[width=0.3\linewidth]{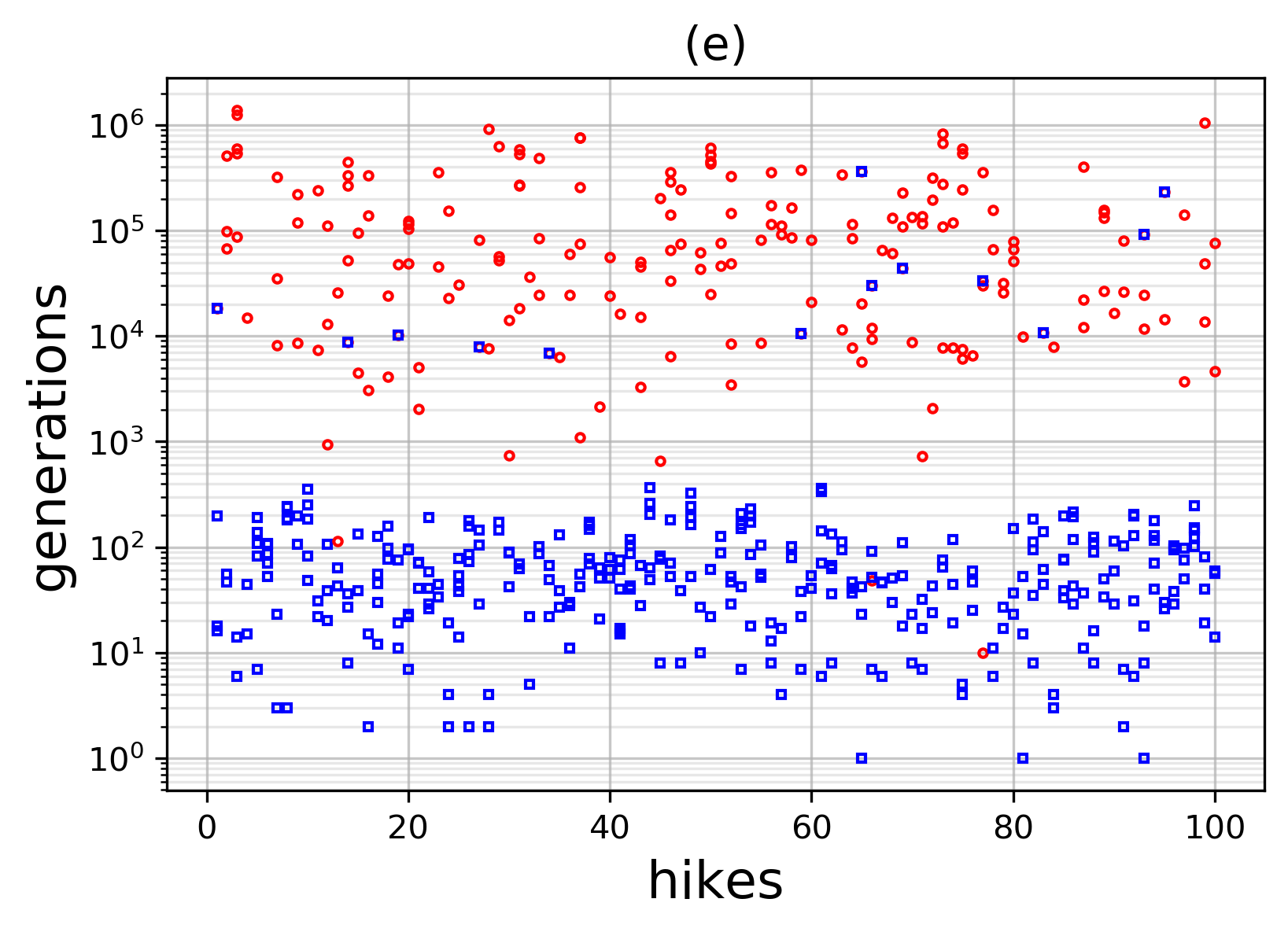}
\includegraphics[width=0.3\linewidth]{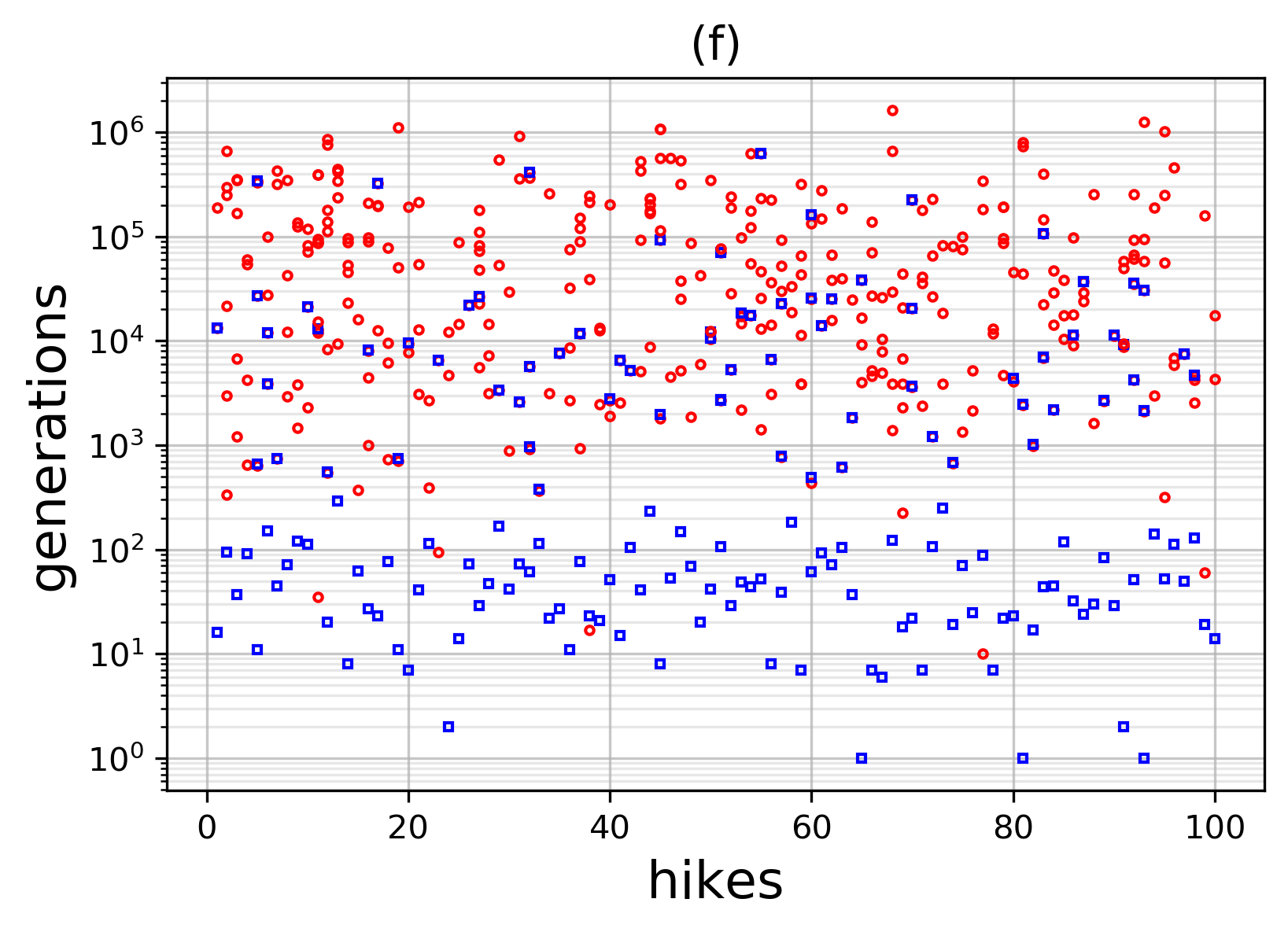}
\caption{{\bf Effect of the degree of epistasis}.
Scatter plots of the mutational events per hike for $N=50$, $p=0.01$, $\tau_{max}=2\times 10^6$ generations. The simulated instances, from ``low'' to ``high'' landscape-ruggedness, are: 
(a) $K=2$, (b) $K=10$, (c) $K=20$, (d) $K=30$, (e) $K=40$, (f) $K=49$.}
\label{fig:hikes2}
\end{center}
\end{figure*}

\paragraph{From low to high ruggedness}
In figure~\ref{fig:hikes2} we compare the behaviour of the evolutionary dynamics for the NK model with different values of the epistatic parameter $K=\{2,10,20,30,40,49\}$ 
($N$ still being fixed to $50$). That is, from a moderately rugged landscape to a full random one. In this case the total number of hikes is $100$ (all starting with a simple genome made of a string of $50$ zeros), $p=0.01$ and $\tau_{max}=2\times10^6$ generations.
In this analysis, inversions occur once every hundred times in average, allowing us to better appreciate the effects of the fitness landscape ruggedness on timescales and 
the duration of the state of stasis due to the epistatic interaction between the $K$ loci. 
For the epistatic interaction values analysed here, it can be seen that during the first stages of the evolutionary process, point mutations always drive evolution, corresponding to an efficient initial exploration of accessible paths (by ``efficient'' we mean that point mutations can be easily and rapidly explored as there are only $N=50$ different possibilities). This period lasts roughly from $ 10 ^ 2 $ to $ 10 ^ 3 $ generations, even though this is less differentiated for the full random landscape ($ K = 49$). 
Then, for $K = 2$, $10$ and $20$, we can identify a gap of evolutionary activity (the stasis-like period) that lasts approximately up to $ 10 ^ 4 $ generations. Some mutations occur during this period, but most of them were preceded by inversions that took place early in the evolutionary process.
Then, from $ K = $ 20, the mutations are more scattered and the stasis-like period tends to vanish. 
When $K$ is large, the number of local fitness peaks increases; therefore, even when genomes are able to escape their current peak, it is likely they have only found a different local optima rather than a global optima. Therefore, since inversions only release genomes from local optima rather than finding global optima, such inversion mutations are less evolutionarily efficient as $K$ increases.
%
%
$K$ also modulates the number of mutational events that occurs during an evolutionary burst (including the initial one), less mutational events being fixed when $K$ increases.

\begin{figure*}[!htp]
\begin{center}
\includegraphics[width=0.3\linewidth]{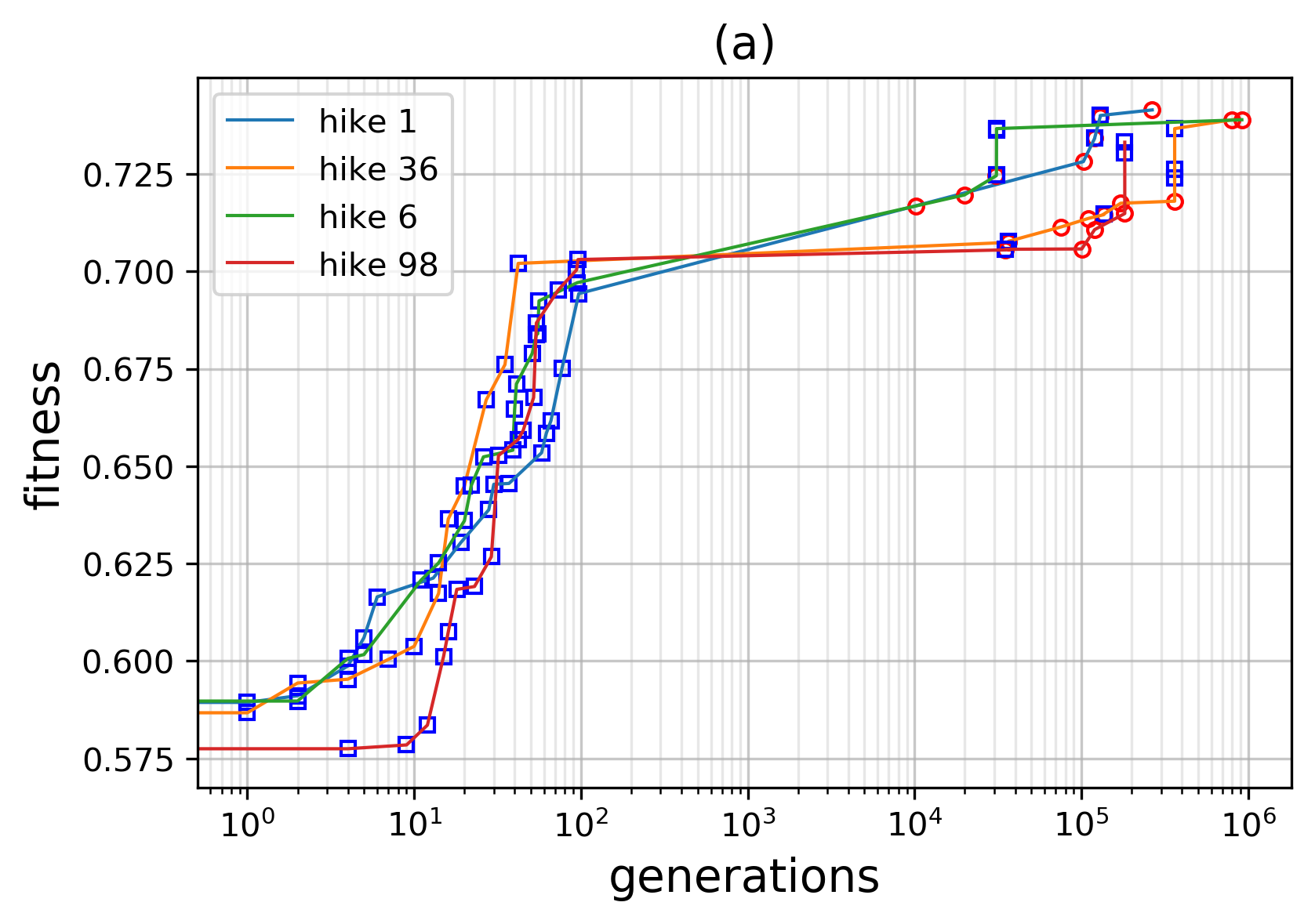}
\includegraphics[width=0.3\linewidth]{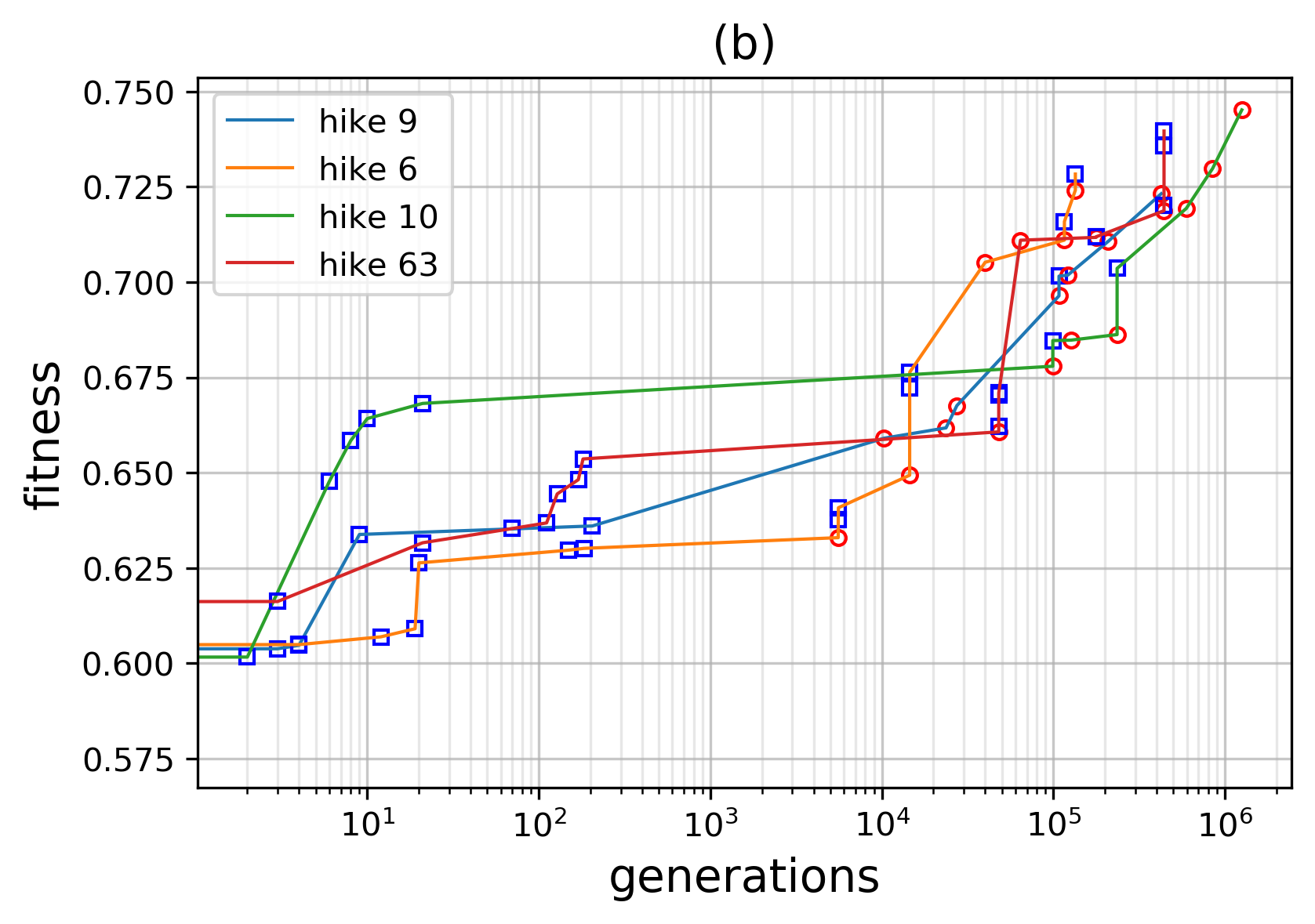}
\includegraphics[width=0.3\linewidth]{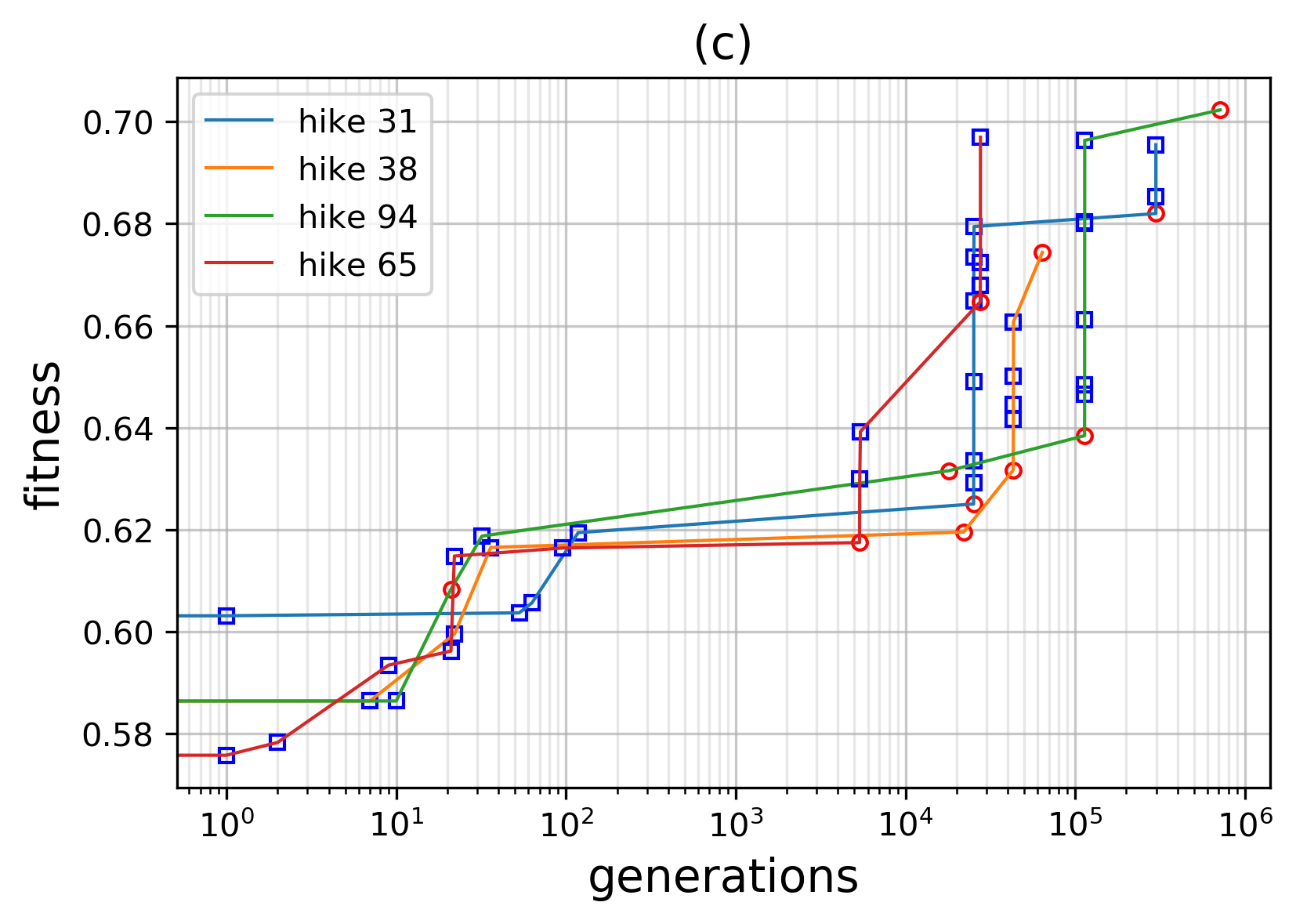}
\includegraphics[width=0.3\linewidth]{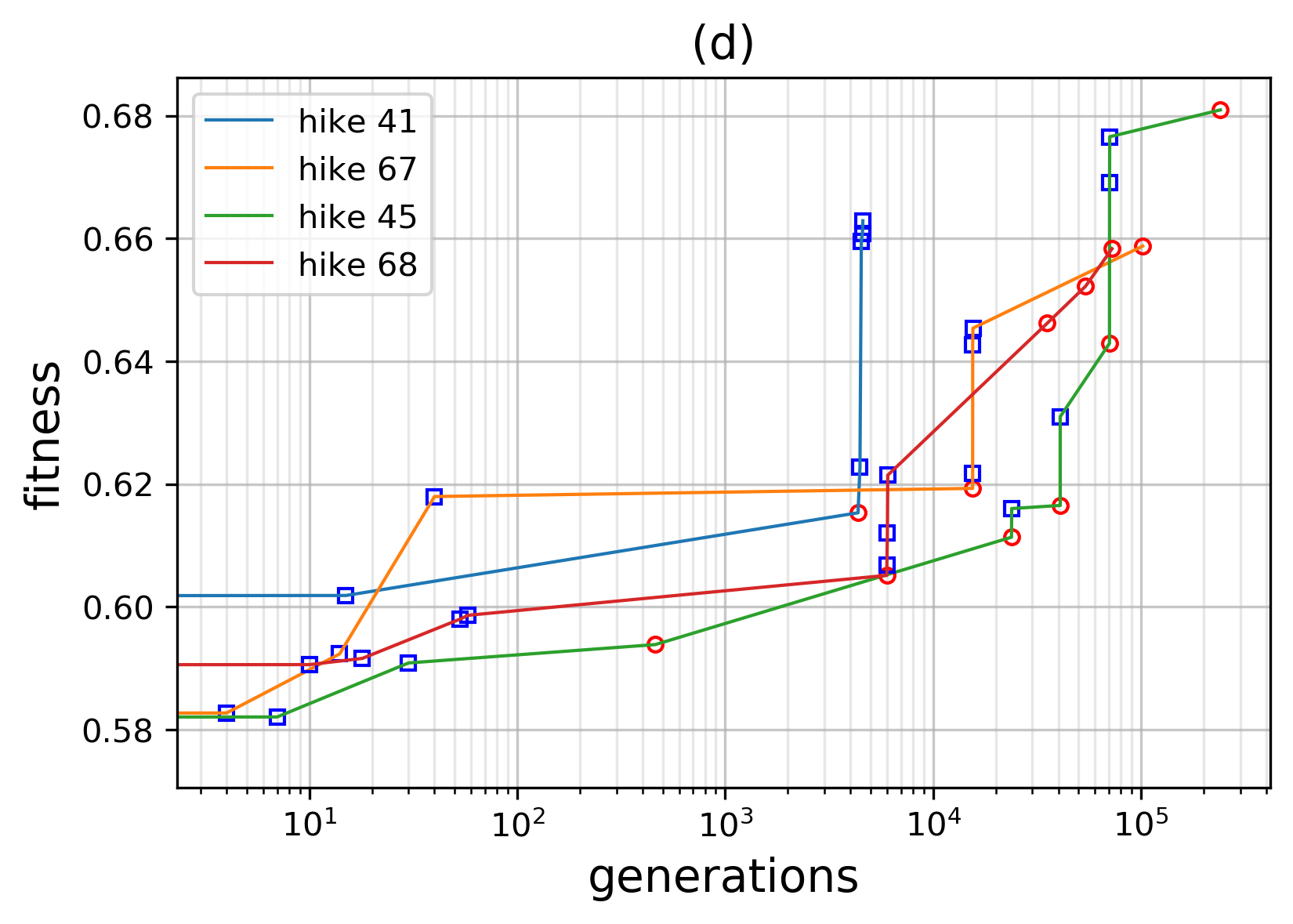}
\includegraphics[width=0.3\linewidth]{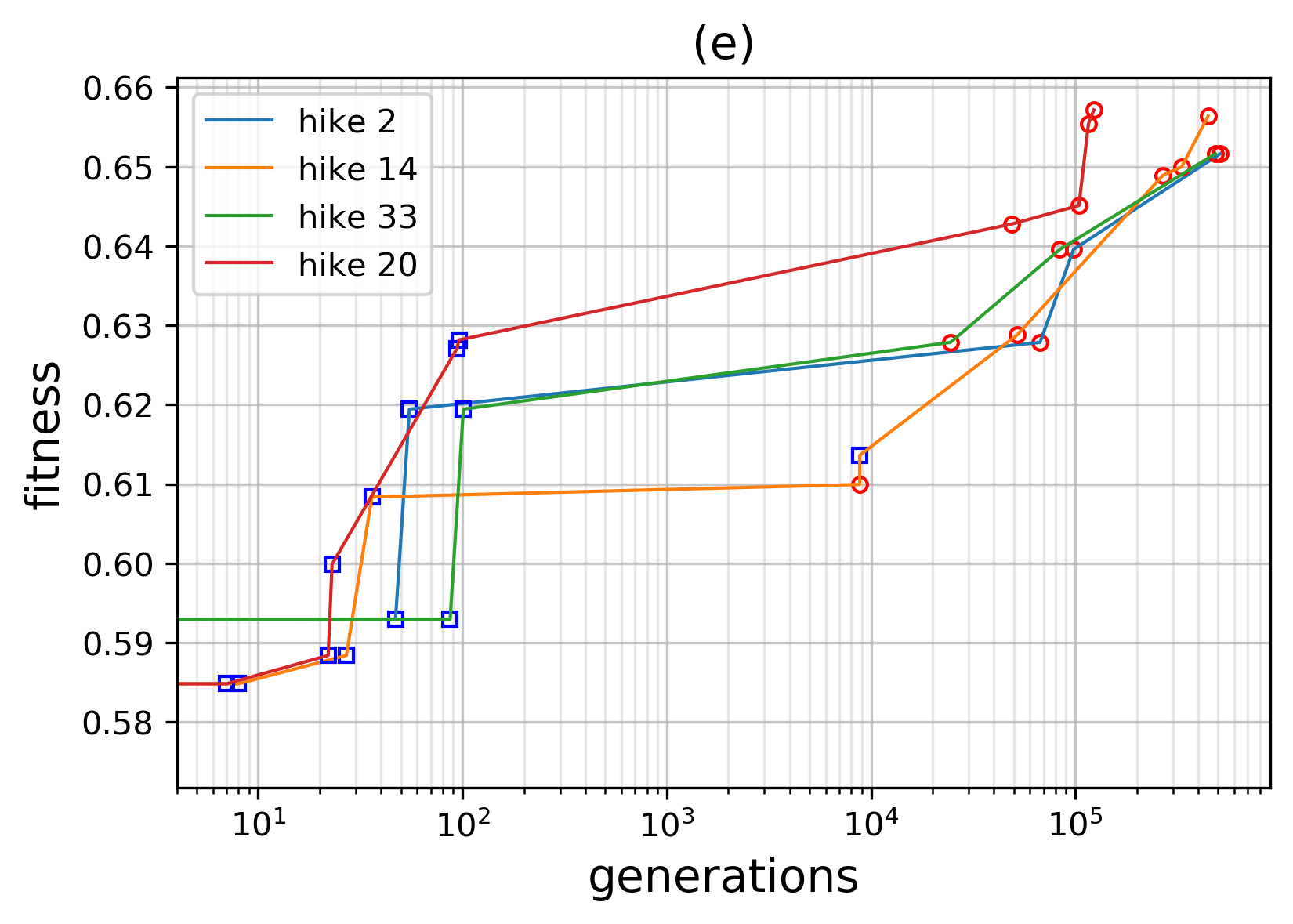}
\includegraphics[width=0.3\linewidth]{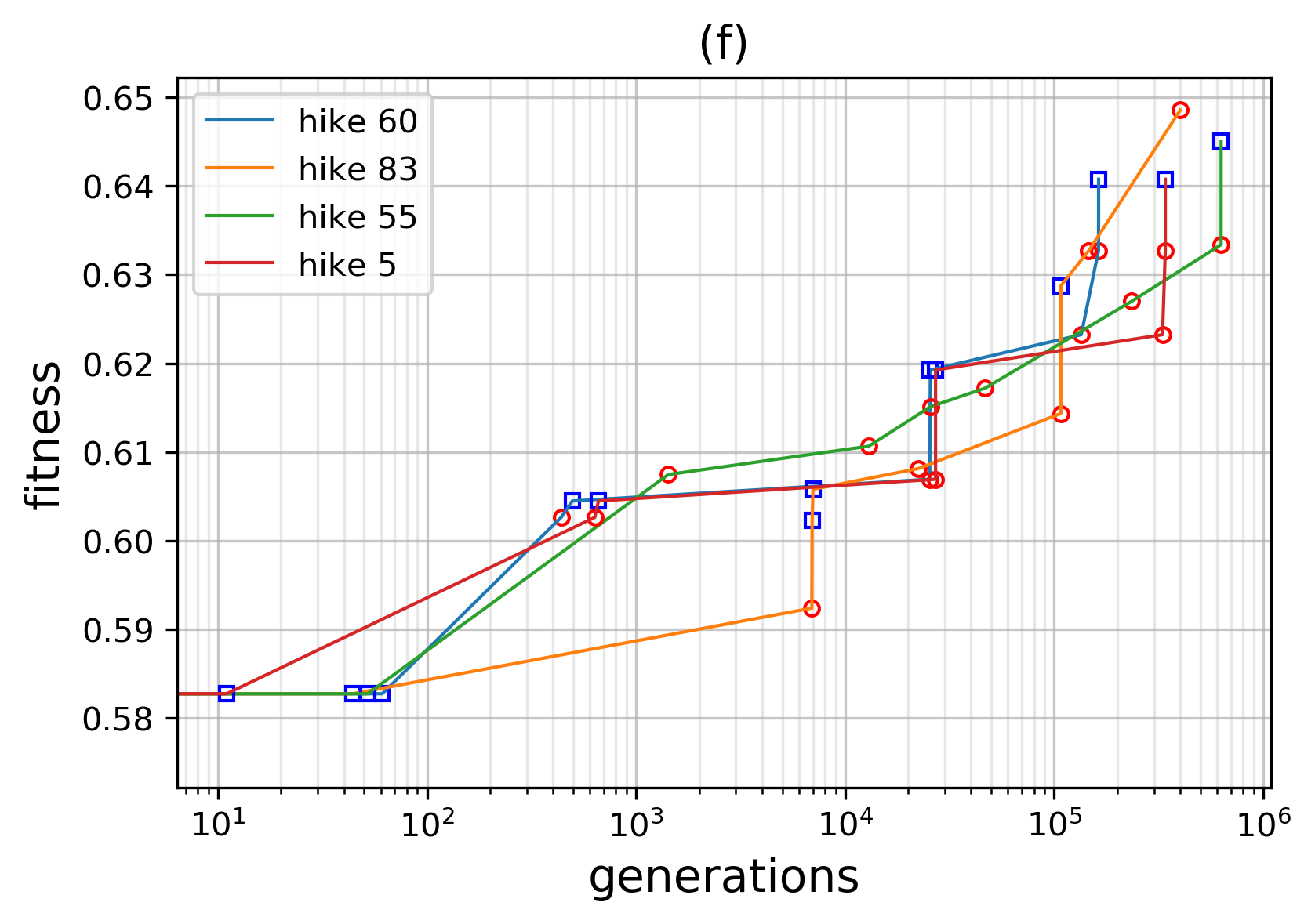}
\caption{{\bf Stasis and burst}. Fitness time-series for a random selection of hikes 
picked from the simulations reported in fig.~\ref{fig:hikes2}, showing the stasis- and burst-like events: 
(a) $K=2$, (b) $K=10$, (c) $K=20$, (d) $K=30$, (e) $K=40$, (f) $K=49$.}
\label{fig:fitness}
\end{center}
\end{figure*}

\paragraph{Examples of time scale dependence with epistasis}
Finally, in figure~\ref{fig:fitness} we present a random selection, from the experiments presented above, of those events that can be identified as possible evolutionary bursts. Here we can see more clearly how inversion mutational events contribute to escaping local fitness peaks, especially in the case when $K = 2$, where it is noted that the different hikes quickly get stuck in local fitness peaks: as the landscape is smoother point mutational events are fixed more often. Once again we see how the evolutionary dynamics reaches a state of stasis, evolution being boosted again with the occurrence of an inversion after a long period of waiting time.  
For $K = 10$, $20$ and $30$, the magnitudes of the fitness increases are quite large, in comparison to the case when $K=2$, which translates into a highly steep slope, giving the impression that the evolutionary process has not only been re-started but also boosted. For the values of $K = 40$ and $49$, the stasis period has almost completely disappeared, and mutational inversion events occur at any evolutionary time scale. This is easily understood, since the increase in the disorder of the (now many) fitness peaks has an effect on the ``randomization'' of evolutionary steps. When epistasis is high, evolutionary dynamics is more ``turbulent'', meaning that there are no well-differentiated time scales between the types of mutations.
We can also see in the scatter plots of the mutational events (figure~\ref{fig:hikes2}) the progressive ``melt'' of  differentiated time scales, where the duration of the stasis is also appreciated as a function of the degree of epistatic interactions.

\section{Conclusions}
From the empirical analysis of our model, it has been shown 
that complex long- and short-term evolutionary dynamics can be
simulated on rugged fitness landscapes. 
To achieve that, we have extended the well-known family of NK landscapes by including inversion-like chromosome rearrangements 
(in this sense, our work complements somewhat related methods to simulate the NK model beyond bit-flips mutations alone, e.g.~\citep{hadany2003,aguirre2003,ostman2012,gillespiechanging}).
We then studied the effects of epistasis on the evolutionary time scales of digital genomes. 

The inversion-like mutations have been incorporated into the simulation of the evolutionary dynamics of NK digital genomes through the probability $ p $, which controls the relative frequency of occurrence of each kind of mutation. By combining the effect of $p$ and $K$ (level of epistasis), we have been able to reproduce behaviors similar to long waiting periods in evolutionary dynamics -- suggestively evoking evolutionary stasis -- followed by mutational bursts.
Therefore, with this in hand we have been able to model the length of adaptive paths as a composition between two different mutational events, the traditional bit-flip that explores the neighborhood of a short-range domain of the fitness landscape, 
and the effects of genome sequence inversion, akin to long-range exploration~\citep{bagnoli2001,tomassini2016}.
All this, together with the combinatorial aspect of all possible variations of any binary sequence, also determine the orders of magnitude of the time scales. Indeed, combinatorics  imposes its time scale which depends on the size of the binary sequence and on the specifics of the different mutational operations and therefore deserves to be studied in more detail than the results presented here.

At least three main time scales were found in our simulations. The first one, is an exploration mainly driven by bit-flips up to a local maximum of fitness. Then, a stasis-like regime where almost no mutation is fixed. This stationary state can be broken with the fixation of an inversion-like mutation, allowing for the exploration of a new region of the fitness landscape. Once the digital organism ``jumps off'' a local peak, it can reach a new ``hill'' in the fitness landscape.
This opens the possibility of a new exploration, which is again carried out mainly by point mutations that may lead to burst-like events, i.e. the rapid succession of point mutations following an inversion. 
Indeed, in the present model bursts can only be triggered by chromosomal inversions but real biological sequences may undergo a vast repertoire of chromosomal rearrangements (inversions, translocations, duplications...) that deserves future studies.
The system may then alternates between phases of stasis and bursts (triggered by inversion like mutations) up to a ``dead-end'' where all accessible favorable paths have been followed and the evolutionary potential of the (last mutated and selected) binary genome has been exhausted and/or blocked by epistatic interactions. Note that such ``dead-end'' may be relatively quickly reached in our simulations (the size of the mutational repertoire being proportional on the genome size $N$) but that, in longer/real genomes, the number of possible inversions ($N^2$) quickly grows. We suggest that this enables virtually infinite exploration of new possibilities, akin to open-endedness~\citep{banzhaf2016defining}. This is another topic that deserves to be investigated and that is also a subject of debate in the evolutionary biology and artificial life communities.

The fitness time-series reported in this work are examples of the statistics of records increments, and the coexistence of short- and long-range time scales resembles the drift and diffusion jumps phenomenology. That is a Brownian-like random walk including L\'evy flights~\citep{masuda2017}. 
This was also evidenced in the scatter plots of figure~\ref{fig:hikes2} and the transition-like
behaviour towards a more ``turbulent'' evolutionary dynamic modulated through epistasis. 

Finally, our model reproduces an evolutionary pattern alternating short and long time scales, without resorting to a description at the population level. It is often supposed that the discrepancy between short and long evolutionary time scales is due to long-term environmental variations \citep{uyeda2011}. Despite its simplicity, our model shows that such patterns can occur in stable environments through the sole intricacy of different mutation operations having different exploration time scales. The present work highlight that mutations which locally reverse the genome order mimic the process called by \cite{kauffman2000} as ``expanding the adjacent possible''. In our case, this can be translated as ``inversion-like mutation driving digital organisms into a new domain on the fitness landscape'', hence ending possibly long-lasting stasis periods and triggering bursts of bit-flip mutations. In this sense, our model could describe a simple scenario that, nonetheless, can shed new light on punctuated equilibrium and complex interlocking micro- macroevolutionary dynamics.


\footnotesize

\bibliographystyle{apalike}

\end{document}